\documentclass[seceq]{ptptex}

\usepackage{graphicx,bm}

\newcommand{\mybm}[1]{\mbox{\boldmath$#1$}}
\newcommand{\mysw}[1]{\scriptscriptstyle #1}

\title{
Renormalization Group Potential for Quasi-One-Dimensional Correlated Systems%
}

\author{
Ming-Shyang \textsc{Chang}$^{1,2}$, Wei \textsc{Chen}$^{1,3}$ and Hsiu-Hau \textsc{Lin}$^{1,4}$%
}

\inst{
$^1$ Department of Physics, National Tsing-Hua University,
Hsinchu 300, Taiwan\\
$^2$ Department of Physics and Astronomy, University of British Columbia, Vancouver, BC V6T 1Z1, Canada\\
$^{3}$ Department of Physics, University of Florida, Gainesville, FL 32611, U. S. A.\\
$^4$ Physics Division, National Center for Theoretical Sciences, Hsinchu 300, Taiwan}

\abst{
We studied the correlated quasi-one-dimensional systems by one-loop renormalization group techniques in weak coupling. In contrast to conventional g-ology approach, we formulate the theory in terms of bilinear currents and obtain all possible interaction vertices. Furthermore, the one-loop renormalization group equations are derived by operator product expansions of these currents at short length scale. It is rather remarkable that these coupled non-linear equations, after appropriate rescaling, can be casted into potential flows. The existence of what we nicknamed ``RG potential" provides a natural explanation of the emergent symmetry enhancement in ladder systems. Further implications arisen from the RG potential are also discussed at the end.
}

\begin{document}
\maketitle

\section{Introduction}

Quasi-one-dimensional (Q1D) systems have attracted extensive attentions
from both experimental and theoretical aspects\cite{Giamarchi04,Dagotto96}. 
Due to the low dimensionality, strong quantum fluctuations often give rise to
surprising behaviors, which are rather different from our intuitions built in higher dimensions. It is then exciting to explore various exotic phenomena, such as spin-charge separation\cite{Lieb68,Kim96}, unconventional electron pairing\cite{Noack94,Balents96}, spin liquids\cite{Lin98,Tsuchiizu02}, broken time-reversal symmetry\cite{Fjaerestad02,Marston02} and so on, in the correlated Q1D systems\cite{Fabrizio93,White94,Tsunetsugu94,Troyer96,Khveshchenko96,Schulz96,Arrigoni96,Lin97,Emery99,Ledermann00,Ledermann01}. Note that, in addition to the ladder compounds, carbon nanotubes, nanoribbons and quantum wires, after integrating out fluctuations at higher energy, are also described by the Q1D theory. Therefore, not only posting a challenging task for academic curiosity, the understanding of the Q1D systems now becomes crucial as the technology advances to the extremely small nanometer scale.

Since mean-field theory is not appropriate in the presence of strong quantum fluctuations, the ground state properties are studied by combinations of powerful techniques such as exact solution (Bethe Ansatz), renormalization-group (RG) analysis, bosonization and numerical simulations. In this article, we would mainly focus on the RG analysis for Q1D correlated systems and reveal several interesting hidden structures behind the complicated RG flows. However, we would like to emphasize that RG analysis only gives rise to various instabilities of the homogenous Fermi liquid but does not provide a complete description of the interaction-driven phases. A non-perturbative complementary approach like bosonization is necessary to understand the physical properties in low-energy limit. Since the combination of RG analysis and bosonization have become rather standard now, interested readers are directed to several useful reviews in the literature\cite{Giamarchi04}.

In the following, we would give a brief introduction to the current understanding of the ground states in Q1D systems. In some of the ladder compounds, the ground state is an (charge) insulator. Since the charge sector is frozen out, the ground-state phase diagram is greatly simplified and can be determined solely from the remaining spin degrees of freedom\cite{Affleck94}. Early theoretical studies for Heisenberg ladders revealed an interesting odd-even effect\cite{Schulz86,Dagotto92,Rice93,White94}. If the number of legs $N$ is even, the system is expected to be a spin liquid with a finite gap to all spin excitations. On the other hand, if $N$ is odd, the ground state has quasi-long-range antiferromagnetic order and the low lying excitations are gapless and belong to the universality class of the single Heisenberg chain.

The subtle interplay between charge and spin fluctuations bring out rich physics and exotic surprises. For instance, it has been speculated for a long time that a doped spin liquid may give rise to unconventional superconductivity. Let's takle the simplest and well-studied two-leg ladder\cite{Dagotto92,Balents96,Lin98,Noack94,Noack95,Sano96,Troyer96,Vojta99,Kim99,Oitmaa99} as an example. Without doping, there is one electron per site on average (usually referred as ``half-filled''). Because of the Coulomb repulsion, charge excitations acquire a gap which makes the two-leg ladder a Mott's insulator. Besides, due to the tendency for singlet bond formation across the rungs of the ladder, spin-liquid behavior is 
expected. Both numerical and analytical approaches support that the 
ground state at half filling is in the Mott-insulating spin-liquid 
phase \cite{Balents96,Lin98,Noack94,Noack95}. Upon doping, the charge gap vanishes, these preformed pairs give rise to the quasi-long-range superconductivity order. It is rather interesting that the pairing symmetry of the Cooper pairs is the unconventional $d$-wave, in contract to the $s$-wave symmetry for phonon mediated superconductivity. This provides a concrete example for unconventional pairings and thus superconductivity in electron gas in the presence of repulsive interactions.

Although our understanding of the strictly one-dimensional systems is greatly
benefitted from exact solutions, it is often found that the Q1D systems are not soluble analytically. Furthermore, since the hopping along the transverse direction is relevant, the physics phenomena for Q1D systems can be dramatically different from the strictly 1D systems. Therefore, the most reliable approach to clarify the competitions among various ground states is the renomalization group (RG) analysis. Except for some exactly soluble models, the phase diagram of Q1D systems is obtained by first spotting the instabilities in RG flows and then analyzed with non-perturbative bosonization techniques. For $N$-leg ladders, one typically needs to solve order $N^2$ coupled non-linear first-order differential equations to obtain the flows. Since the number of allowed interactions are large, the derivation of the flow equations for all couplings under RG transformations becomes formidable. On the other hand, both numerical and analytical approaches seem to indicate rather simple phase diagrams for generic Q1D systems.

As one of the main points in this article, we will address the issue why the phase diagrams for quasi-one-dimensional systems are rather simple, while the renormalization group equations behind the scene are non-linear and messy looking. The puzzle is answered in two steps -- we first demonstrate that the complicated coupled flow equations are simply described by a potential $V(h_i)$, in an appropriate basis for the interaction couplings $h_i$. The renormalization-group potential is explicitly constructed by introducing the Majorana fermion representation. The existence of the potential prevents chaotic behaviors and other exotic possibilities such as limit cycles. Once the potential is obtained, the ultimate fate of the flows are described by a special set of fixed-ray solutions and the phase diagram is determined by Abelian bosonization.  Extension to strong coupling regime and comparison with the Zamolodchikov c-theorem will also be discussed.

The simplicity of the phase diagram can be partially understood by the widely used scaling Ansatz of the couplings in weak coupling\cite{Balents96,Lin97,Lin98},
\begin{eqnarray}
g_{i}(l) \simeq \frac{G_{i}}{(l_{d}-l)} \ll 1,
\label{Scaling}
\end{eqnarray}
where $G_{i}$ are order one constants satisfying the non-linear algebraic constraints (discussed later) and $l_{d}$ is the divergent length scale where the flows enter the strong coupling regime. The Ansatz was motivated by the numerical observation that the ratios of renormalized couplings reach constant, as long as the bare interaction strength is weak enough. However, it is still puzzling why the phase diagrams, generated by many coupled non-linear flow equations, do not reflect the same level of complexity. In fact, for a complicated system with many couplings, the coupled non-linear differential equations are likely to produce chaotic flows generically. Even if the flow is not chaotic, it might as well rest on limit cycles. This peculiar possibility for quantum systems was addressed by Wilson and collaborators in a recent
paper\cite{Glazek02}. 

So the question remains: Why are the phase diagram and the RG
flows so simple (without chaotic flow or limit cycle) in Q1D
systems? We found the question can be answered by combining the weak-coupling RG analysis together with the non-perturbative Abelian bosonization technique. Note that the phases of the correlated ground state often lie in the strong coupling, so weak-coupling RG alone can not pin down the phase diagram. It is the powerful combination of the perturbative RG analysis and the non-perturbative bosonization technique which deliver the desired answer here. Note that, in the low-energy limit, the Q1D systems involve $N_f$ flavors of interacting fermions with {\em different} Fermi velocities, where $N_f$ is the number of conducting bands. Although constrained by various symmetries, the number of allowed interactions is tremendously large as $N_f$ increases. In general, the RG equations to the lowest order are already very complicated - not to mention solving them analytically.

However, quite to our surprises, we found that, at one-loop level, the RG flows can be derived from a potential, i.e. the coupled non-linear flow equations can be cast into this elegant form by an appropriate choice of coupling basis,
\begin{equation}
\frac{dh_i}{dl} = - \frac{\partial V(h_{j})}{\partial h_i},
\label{PotentialFlow}
\end{equation}
where $V(h_{j})$ is the RG potential. We emphasize that this is only possible after a unique transformation of the couplings, $h_i(l) = L_{ij} g_j(l)$ (up to a trivial overall factor for all couplings), where $L_{ij}$ is some constant matrix.
The existence of the potential, which requires the coefficients in the RG equations to satisfy special constraints, also provides a self-consistent check on the RG equations derived by other approaches.

The flows of Eq.~(\ref{PotentialFlow}) in the multi-dimensional coupling space can be viewed as the trajectories of a strongly overdamped particle moving in a conservative potential $V(h_i)$. Note that the change of the potential
$V(h_i)$ along the trajectory is always decreasing,
\begin{equation}
\frac{dV}{dl} = \frac{\partial V}{\partial h_i} \frac{dh_i}{dl} =
- \left(\frac{dh_i}{dl}\right)^2 \leq 0,
\end{equation}
where summation over the index $i$ is implicitly implied. Therefore, it is obvious that the function $V(h_{i})$ never increases along the trajectory and is only stationary at the fixed points where $dh_i/dl=0$. Thus, the RG flows have a simple geometric interpretation as the trajectory of an overdamped particle searching for potential minimum in the multi-dimensional coupling space.

This simple geometric picture rules out the possibilities of chaos and
the exotic limit cycles in Q1D systems. The ultimate fate of the flows would either rest on the fixed points or follow along the  ``valleys/ridges" of the
potential profile\cite{Konik02} to strong coupling. Since there is only one trivial fixed point at one-loop order, most of the time, the flows run away from the non-interacting fixed point. Starting from weak enough bare couplings, the ultimate fate of the flows is dictated by the asymptotes of the ``valleys/ridges" of the potential profile. It provides the natural explanation why the ratios of the renormalized couplings reach constant in numerics. That is to say, the existence of RG potential implies that the ultimate fate of RG flows must take the scaling form described in Eq.~(\ref{Scaling}). Detail properties of these asymptotes, referred as fixed rays, will be discussed in later section. 

Since the ultimate fate of RG flows is described by the simple Ansatz in Eq.~(\ref{Scaling}), the specific ratios of couplings simplify the effective Hamiltonian a lot. Making use of the Abelian bosonization, one can determine which sector acquires a gap, triggered from the weak-coupling instability. The phase of the ground state is then determined by watching which fixed ray (asymptote) the flows go closer to. Because there are only limited solutions of the fixed rays, the phase diagram is rather simple. Therefore, by combining the powerful techniques of weak-coupling RG and Abelian bosonization, we pin down the reason behind the simple-looking phase diagram out of the messy non-linear flow equations.

In fact, the combination of weak-coupling RG and Abelian bosonization goes beyond the usual mean-field analysis and is crucially important when there are more than one competing orders\cite{Lauchli04}. For instance, L\"auchli, Honerkamp and Rice recently studied the so-called ``d-Mott" phase in one and two dimensions,  where antiferromagnetic, stagger-flux and d-wave pairing fluctuations compete with each other simultaneously. The conclusion drawn from the numerical density-matrix RG in strong coupling agrees rather well with predictions made from one-loop analysis in weak coupling. This lends support to the powerful combination of weak-coupling RG and Abelian bosonization approach for strongly correlated systems. Since the method of bosonizing the fixed rays is already developed in previous papers\cite{Fabrizio93,Balents96,Lin97,Lin98}, we would concentrate on the novel existence of RG potential in this paper. In particular, we would construct the RG potential explicitly.

The existence of RG potential and its elegant geometric interpretation are also related to the mysterious emergence of symmetry enhancement in ladder systems in low-energy limit. Let us return to the simple example of the half-filled two-leg ladder again. In general, one expects that charge and spin gaps could be rather different. Surprisingly, a complete degeneracy of charge, spin and single particle gaps emerges in the asymptotic weak coupling limit. 
Based on the one-loop renormalization group (RG) analysis, the effective low-energy theory for the two-leg ladder is described by the exactly soluble Gross-Neveu model with a global SO(8) symmetry. Other than the exact degeneracy of excitation gaps, many interesting results can be drawn from the Bethe Ansatz solution. It turns out the emergence of the enhanced symmetry is directly tied up with the ``valleys/ridges" of the RG potential profile in two-leg ladder. One can show analytically that these ``valleys/ridges" in potential profile correspond to asymptotic flows with fixed ratios between all couplings that give rise to the enhanced symmetry. 

There have been critics \cite{Azaria98,Emery99} about the unexpected emergence of the enhanced SO(8) symmetry, questioning the stability of the symmetry under generic perturbations and the limitation of the RG equations derived in weak coupling. It was argued that the emergent symmetry derived by perturbative calculations is fragile because the system eventually flows toward the fixed point in strong coupling. Both concerns can be addressed by performing a complete analytical study for the stability in the vicinity of the symmetric phases. It should be emphasized that this emergent symmetry enhancement in weak coupling does {\it not} implies that the fixed point in strong coupling is SO(8) symmetric at all. On the contrary, the symmetry reflects the local topology of RG flows near the trivial Fermi liquid fixed point which can be safely described by perturbative calculations. Indeed it is rather obvious that, when the interaction $U$ is much larger than the bandwidth $t$, the charge gap 
is much larger than the spin gap. Thus, the symmetry in the strong coupling is completely destroyed. As to the stability of the symmetry under generic perturbations, the complete stability check shows that the SO(8) symmetry is robust in weak coupling. Even though some couplings are relevant according to the conventional classification, they do not destroy the symmetry but only give rise to anomalous corrections which scale as $(U/t)^{1/3}$. A new classification of ``relevant'' and ``irrelevant'' perturbations is necessary here.

The rest of the article is organized as following. In Sec. 2, we present the model for correlated quasi-1D systems and derive the effective theory in weak coupling. Making full use of SU(2)$\times$U(1) symmetries, all interaction vertices can be elegantly expressed in terms of various bilinear currents. In Sec. 3, the one-loop RG equations for the correlated quasi-1D systems are derived from the technique of operator product expansions of bilinear currents. In Sec. 4, we used the half-filled two-leg ladder as example to reveal a hidden structure behind the messy RG equations. By appropriate rescaling of all coupling constants, we show that the RG flows can be derived from a single potential. We also demonstrate that the RG potential also exists for the doped ladders. In Sec. 5, we provide a formal proof for the existence of the RG potential by transforming into the so-called Majorona basis. The potential gives an elegant geometric interpretation of destinations of the RG flows. Finally, in Sec. 6, we address the interesting issue about the emergence of enhanced symmetry in ladder systems and its connection to the existence of RG potential. We also present a simple scaling argument near these symmetric rays and argue that anomalous scaling is expected in the vicinity. 

\section{Quasi-1D Ladders}

We start with the simplest Q1D ladders. In tight-binding limit, the Hamiltonian of the $N$-leg ladder, $H=H_{0} + H_{I}$, contains the nearest-neighbor hopping and generic short-range interactions,
\begin{eqnarray}
H_{0}&=& \sum_{x,i,\alpha}\bigg\{
-t [d_{i\alpha }^{\dag }(x+1) d_{i\alpha
}(x) + d_{i\alpha }^{\dag }(x) d_{i\alpha
}(x+1) ]
\nonumber\\
&& \hspace{1cm} -t_{\perp }[ d_{i+1,\alpha }^{\dag }(x) d_{i\alpha }(x) + d_{i,\alpha }^{\dag }(x) d_{i+1\alpha }(x) ]
\bigg\},
\\
H_{I} &=& \sum_{x,x',i,j} n_{i}(x) V_{i,j}(x-x') n_{j}(y),
\end{eqnarray}
where $d_{i}(d_{i}^{\dagger })$ is the annihilation (creation) operator for
the fermion on chain $i$ ($i=1...N$) and $\alpha =\uparrow ,\downarrow $ is the spin index. The parameters $t$ and $t_{\perp }$ denote the hopping amplitudes along the same chain and across the rungs between neighboring chains. The density operator on the ladder is $n_i(x) \equiv \sum_{\alpha} d_{i\alpha}^\dag(x) d_{i\alpha}(x)$. The short-range interaction $V_{i,j}(x-x')$ is taken to be the density-density type here but it can be easily generalized. Note that the method we try to build here is for {\em generic} interaction profile, not limited to the on-site repulsion.

In weak coupling, it is natural to concentrate on the hopping Hamiltonian $H_0$ first, before including the correlation effects from $H_I$ perturbatively. Depending on the boundary condition in the transverse direction, the nearest-neighbor hopping can be diagonalized easily. Since the open boundary condition (OBC) is more natural and relevant to the realistic ladder materials, we will concentrate on this case first. However, we also include the studies on periodic boundary condition (PBC) later for completeness. For OBC, the transverse hopping across the rungs can be brought into diagonal form by the $N \times N$ matrix $S_{jm}$,
\begin{equation}
d_{j\alpha}(x)=\sum_{m}S_{jm}\psi _{m\alpha}(x),\quad \mbox{with} \hspace{4mm} S_{jm}=\sqrt{\frac{2}{N+1}}\sin \left(\frac{\pi jm}{N+1} \right).
\end{equation}
Furthermore, after transforming into the momentum space along the chain direction, the hopping Hamiltonian takes the simple form,
\begin{equation}
H_{0}=\sum_{i,\alpha}\int_{-\pi}^{\pi}\frac{dk}{2\pi}\: 
\epsilon _{i}(k) \psi _{i\alpha }^{\dag }(k)\psi _{i\alpha }(k).
\end{equation}
The single-particle spectrum is $\epsilon_{i}(k)=-2t\cos k -2t_{\perp}\cos (k_{yi})$, where the quantized momentum in transverse direction is $k_{yi}=\pi i/(N+1)$.

For the purpose of deriving effective theory in low-energy limit, it is sufficient to linearize the energy spectrum near the Fermi points that intersect with the chemical potential $\epsilon_i(k_{Fi}) = \mu$. This implies the lattice operator $\psi_{i\alpha}(x)$ can be decomposed into a pair of chiral fields at each Fermi points,
\begin{equation}
\psi _{i\alpha }(x)\;\sim \;\psi _{Ri\alpha }(x)\;e^{ik_{Fi}x}\;+\;\psi
_{Li\alpha }(x)\;e^{-ik_{Fi}x}.
\end{equation}
Since the fast-oscillatory phase factors near the Fermi points are pulled out already, the chiral fields $\psi_{\mysw{P}i\alpha}(x)$ with $P=R,L$ are smooth varying operators. Finally, the hopping Hamiltonian can be expressed in terms of these chiral fields with linear dispersion,
\begin{equation}
H_{0}=\sum_{i,\alpha }\int dx\ v_{i}\left[ -\psi _{Ri\alpha }^{\dagger
}(x)\;i\partial _{x}\ \psi _{Ri\alpha }(x)+\psi _{Li\alpha }^{\dagger
}(x)\;i\partial _{x}\ \psi _{Li\alpha }(x)\right]
\end{equation}
where $v_{i}=2t\sin (k_{Fi})$ is the Fermi velocity of the $i$-th band.

Before we dive into the details of how to derive all possible interactions, we would like to comment on the validity of the weak-coupling theory first. To address this issue, let's write down the partition function for the correlated Q1D ladder first,
\begin{eqnarray*}
Z\;=Tr\;e^{-\beta H}=\int D[\overline{\psi} \psi] e^{-S}.
\end{eqnarray*}
The Euclidean action can be derived from the path integral,
\begin{equation}
S=\int_{0}^{\beta }d\tau \left[ \sum_{x,i,\alpha}\overline{\psi}
_{i\alpha }(x)\;\partial _{\tau } \psi _{i\alpha }(x)+H\right].
\end{equation}
The validity of chiral-field decomposition relies on how the interaction strength is renormalized after integrating out the modes that are far from the Fermi points. To be more precise, the momentum regimes $\mid p-k_{Fi}\mid >\Lambda $ and $\mid p+k_{Fi}\mid >\Lambda$ that are outside the vicinity of the Fermi points, are integrated out in the path integral. The standard perturbation theory gives the renormalized interaction strength
\begin{equation}
U_{R}\simeq U\left[ 1+ \mbox{const} \times \frac{U}{t}\ln \left(\frac{k_{Fi}}{\Lambda}\right)\right] .
\end{equation}
Therefore, as long as $(U/t) \ll 1/\ln (\frac{k_{Fi}}{\Lambda }) \sim {\cal O}(1)$, the one-step RG from the whole Brillouin zone down to the cutoff $\Lambda$ does not give significant renormalization and the perturbative approach is reliable.

We are now ready to search for all possible interaction vertices and write them down in terms of the chiral fields. Note that, at half filling, there are special constraints between Fermi momenta, $k_{Fi} + k_{F\hat{i}}=\pi$, where $\hat{i} = N+1-i$. These constraints give rise to additional vertices known as umklapp interactions. In addition to the U(1) $\times$ SU(2) symmetry corresponding to charge and spin conservation, these terms must be preserved by charge conjugation, time reversal, parity, and (lattice) translations. The most general particle-conserving 4-point
vertex takes the form,
\begin{eqnarray}
{\cal H}_{int}=\sum_{P_a, i_a} G[P_a,i_a]
\psi^{\dag}_{\mysw{P}_1i_1}(x)\psi^{\dag}_{\mysw{P}_2 i_2}(x)
\psi^{}_{\mysw{P}_3 i_3}(x)\psi^{}_{\mysw{P}_4 i_4}(x).
\end{eqnarray}
Here the spin indices are suppressed for notational clarity. We will include them back explicitly later.

Various discrete symmetries, such as charge conjugation, time reversal, parity and so on, give constraints on $P_a, i_a$. However, the strongest limitation arises from conservation of lattice momentum, imposing constraints on the allowed band indices $(P_{a}, i_{a})$
\begin{eqnarray}
-P_1 k_{Fi_{1}}-P_2 k_{Fi_{2}}
+P_3 k_{Fi_{3}}+P_4 k_{Fi_{4}}=2\pi n_{x}.
\label{Kx_cons}
\end{eqnarray}
Although we deal with OBC here, it is important to remind the readers that the transverse momentum is also conserved when PBC is considered. Thus, there are additional constraints associated with the translational invariance in the transverse direction,
\begin{eqnarray}
-P_1i_1-P_2i_2+P_3i_3+P_4i_4=N n_{y} \:\:
\mbox{(PBC only!!)}. 
\label{Ky_cons}
\end{eqnarray}
The remaining work is to find out all possible solutions for $(P_{a}, 
i_{a})$ in Eqs.~(\ref{Kx_cons})-(\ref{Ky_cons}). When searching for the
solutions, one must keep in mind that a solution with the same chirality for all $P_a$ does not cause any instability in the ground state. Instead, it only gives rise to renormalization of Fermi velocities and, therefore, can be safely ignored for the derivation of one-loop RG equations. Since they are irrelevant to the phase diagram of ground states, we will ignore them in the following discussions.

\subsection{Open Boundary Conditions}

For OBC, only momenta in the $k_{x}$ direction is conserved. 
Therfore, only Eq.~(\ref{Kx_cons}) needs to be studied. To pin down all possible solution is literally impossible. However, one notice that we can classify the solutions into two categories: the major vertices that always exist for generic parameters $t, t_{\perp}, n, ...$ and the minor vertices that only show up when the parameters are specifically fine-tuned. Since the phase space of the minor vertices are negligible in comparison with the major ones, it is reasonable to drop them. However, one should keep in mind that the approximation may break down if the system under consideration has peculiar shape of Fermi surface that greatly enhances the minor vertices.
The reason why the major vertices appear in the whole parameter space implies the momentum conservation in Eq.~(\ref{Kx_cons}) must be satisfied {\em trivially}. Thus, we look for linear combinations of some trivial identities for half-filled Q1D ladders. For OBC, these identities are rather simple,
\begin{eqnarray}
k_{Fi}-k_{Fi} &=& 0,
\label{identity1}
\\
k_{Fi}+k_{F\hat{i}}&=&\pi,
\label{identity2}
\end{eqnarray} 
where $\hat{i} \equiv (N+1)-i$ is the band index which has the same Fermi 
velocity as the $i$-th band. The first identity is indeed trivial and the second one is protected by the particle-hole symmetry at half-filling.

Vertices can be classified by the integer $n_{x}$ in Eq.~(\ref{Kx_cons}).
Since $k_{Fi} < \pi$, $n_{x}$ can take values, $0, \pm1$. For $n_{x}=0$, we can construct the linear combination of Eq.~(\ref{identity1}),
\begin{equation}
k_{Fi}-k_{Fi}+k_{Fj}-k_{Fj} = 0.
\end{equation}
Compared with Eq.~(\ref{Kx_cons}), it is straightforward to spot two
sets of solutions. The first set comprises forward scattering that satisfies
\begin{eqnarray}
  (P_1, P_2)&=&(P_3, P_4)
  \nonumber\\
  (i_1,i_2) &=&(i_3,i_4), 
  \hspace{1cm} {\rm (forward),}
  \label{forward_cond}
\end{eqnarray}
The second set is the Cooper (or backward) scattering,
\begin{eqnarray}
P_1=\overline{P}_2, &\qquad&P_3=\overline{P}_4;
\nonumber\\
i_1=i_2, &\qquad&i_3=i_4,
\hspace{1cm}{\rm (Cooper).}
\label{Cooper_cond}
\end{eqnarray}
In Eqs.~(\ref{forward_cond})-(\ref{Cooper_cond}), and in the remainder of
the paper, $\overline{P} \equiv -P$ and $(x_1,x_2)=(x_3, x_4)$
indicates pairwise equality, i.e. either $x_1=x_3$, $x_2=x_4$, or
$x_1=x_4$, $x_2=x_3$.  The two possible solutions for forward
scattering actually describe the same vertices, up to a sign from the
fermion ordering.  Referring to the 2D Brillouin zone, one sees that
forward scattering preserves the particle number in each band
separately, i.e. electrons are annihilated/created within the same band. For Cooper scattering, pairs of electrons are annihilated in one band (for instance, the $i_1$-th band) and then scattered into anotehr band (for instance the $i_3$-th band).

There is another set of solution for $n_x = 0$. Making use of the linear combination of Eq.~(\ref{identity2})
\begin{equation}
k_{Fi}+k_{F\hat{i}}-k_{Fj}-k_{F\hat{j}}=0,
\end{equation}
we found another set of vertices which involve {\em four} different bands. In contrast to the forward and Cooper scatterings that only involve two different bands, this strange set of solutions requires
\begin{eqnarray}
(P_1, P_2) &=& (P_3, P_4),
\quad P_1=\overline{P}_2, P_3=\overline{P}_4;
\nonumber\\ 
(i_1,i_2) &= & (\hat{i}_{4}, \hat{i}_{3}).
\label{Nest_cond}
\end{eqnarray}
Note that the existence of this strange set of vertices crucially relies on the nested Fermi surface at half filling. However, one should not confuse it with the umklapp interactions (where $n_x \neq 0$) because the momenta in these strange vertices are exactly conserved. 

In addition to the three set of vertices for $n_x=0$. We now turn our attentions to the so-called umklapp interactions with non-vanishing $n_x$. For $n_{x}= \pm 1$, the only useful combination from  Eq.~(\ref{identity2}) is,
\begin{equation}
k_{Fi}+k_{F\hat{i}}+k_{Fj}+k_{F\hat{j}}=2\pi.
\end{equation}
Compared with Eq.~(\ref{Kx_cons}), it gives two sets of solutions. The first set is the usual umklapp interaction that requires
\begin{eqnarray}
P_{1}=P_{2}&=&\overline{P}_{3}=\overline{P}_{4};
\nonumber\\
(i_{1}, i_{2})&=&(\hat{i}_{3}, \hat{i}_{4}).
\end{eqnarray}
This set of vertices describes the scattering process for a pair of electrons from bands $i$ and $j$ with one chirality to their particle-hole symmetric partners, bands $\hat{i}$ and $\hat{j}$, with opposite chirality. The change of momentum associated with this kind of vertices is not zero, $\Delta P = \pm (k_{Fi}+k_{Fj}+k_{F\hat{i}}+k_{F\hat{j}}) = \pm 2\pi$ and thus belong to the category of umklapp interactions.

The other set of solutions requires
\begin{eqnarray}
P_{1}=P_{2}&=&\overline{P}_{3}=\overline{P}_{4};
\nonumber\\
i_{1}=\hat{i}_{2}&& i_{3}=\hat{i}_{4}.
\end{eqnarray}
This set of vertices describes the scattering process for a pair of electrons from particle-hole symmetric bands $i$ and $\hat{i}$ with one chirality to another pair of bands $j$ and $\hat{j}$ with opposite chirality. Again, this kind of vertices falls into the category of umklapp interaction because the change of momentum associated with this kind of vertices, $\Delta P = \pm (k_{Fi}+k_{F\hat{i}}+k_{Fj}+k_{F\hat{j}}) = \pm 2\pi$, is not zero.

Collect all the results we got so far, the allowed interactions consist of three kinds of momentum-conserving vertices and two kinds of umklapp interactions. It is possible to write down the
Hamiltonian density ${\cal H}_{int}={\cal H}^{(1)}_{int}
+{\cal H}^{(2)}_{int}$ in terms of the chiral fields,
\begin{eqnarray}
 {\cal H}^{(1)}_{int} &=& \sum_{P,i,j} \bigg\{
  C[P,i,j] \psi^{\dag}_{\mysw{P}i}\psi^{\vphantom{\mysw{P}}}_{\mysw{P}j}
  \psi^{\dag}_{\overline{\mysw{P}}i}\psi^{\vphantom{\mysw{P}}}_{\overline{\mysw{P}}j}
  \nonumber\\
  && +\:\: F[P,i,j] \psi^{\dag}_{\mysw{P}i}\psi^{\vphantom{\mysw{P}}}_{\mysw{P}i}
  \psi^{\dag}_{\overline{\mysw{P}}j}\psi^{\vphantom{P}}_{\overline{\mysw{P}}j} 
  \nonumber\\
  && +\:\: S[P,i,j] \psi^{\dag}_{\mysw{P}i}\psi^{\vphantom{P}}_{\mysw{P}j}
  \psi^{\dag}_{\overline{\mysw{P}}\hat{j}}
  \psi^{\vphantom{P}}_{\overline{\mysw{P}}\hat{i}}
  \bigg\},
\label{realspace1}
\end{eqnarray}
\begin{eqnarray}
{\cal H}^{(2)}_{int}&=&\sum_{P,i,j} \bigg\{
U[P,i,j] \psi^{\dag}_{\mysw{P}i}\psi^{\dag}_{\mysw{P}j}
\psi^{\vphantom{P}}_{\overline{\mysw{P}}\hat{i}}
\psi^{\vphantom{P}}_{\overline{\mysw{P}}\hat{j}}
\nonumber\\
&&  + \:\: W[P,i,j] \psi^{\dag}_{\mysw{P}i} \psi^{\dag}_{\mysw{P}\hat{i}}
\psi^{\vphantom{P}}_{\overline{\mysw{P}}j}
\psi^{\vphantom{P}}_{\overline{\mysw{P}}\hat{j}} 
\bigg\}.
\label{realspace2}
\end{eqnarray}
Note that ${\cal H}^{(1)}_{int}$ contains vertices which conserve momentum, 
while ${\cal H}^{(2)}_{int}$ denotes the umklapp interactions in the $k_{x}$ direction.

Now we need to include the spin indices. If we follow the conventional g-ology approach, it is rather messy to write down all vertices explicitly. To make the full use of the rotational symmetry in spin sector, it is very helpful to express the four-fermion interactions in terms of SU(2) bilinear currents. We introduce the SU(2) scalar and vector currents,
\begin{eqnarray}
 J_{\mysw{P}ij} = \frac12 \psi^{\dag}_{\mysw{P}i\alpha} \psi^{\vphantom\dag}_{\mysw{P}j\alpha}, 
 &\qquad&
 \mybm{J}_{\mysw{P}ij}=\frac12 \;\psi^{\dag}_{\mysw{P}i\alpha} 
 \mybm{\sigma}_{\alpha\beta} \psi^{\vphantom\dag}_{\mysw{P}j\beta},
 \label{J_def}
 \\
I_{\mysw{P}ij} = \frac12 \psi_{\mysw{P}i\alpha} \epsilon_{\alpha\beta} \psi_{\mysw{P}j\beta},
&\qquad&
\mybm{I}_{\mysw{P}ij}=\frac12 \;\psi_{\mysw{P}i\alpha} (\epsilon\mybm{\sigma})_{\alpha\beta} \psi_{\mysw{P}j\beta},
\label{I_def}
\end{eqnarray}
where $\mybm{\sigma}$ is Pauli matrices and $\epsilon$ 
is the Levi-Civita antisymmetric tensor with the convention
$\epsilon_{12}=-\epsilon_{21}=1$. To regularize the composite 
operators in Eqs.~(\ref{J_def})-(\ref{I_def}), the defined currents are 
already normal ordered (although we do not indicate the normal ordering explicitly). As is clear, $J_{ij}$, $I_{ij}$ are scalars under SU(2) 
transformation, while $\mybm{J}_{ij}$, $\mybm{I}_{ij}$ are vector 
currents because of the Pauli matrices $\mybm{\sigma}$. Furthermore, due to Fermi statistics, it is straightforwad to show the scalar current $I_{ij}$ is symmetrical, while the vector one $\mybm{I}_{ij}$ 
is anti-symmetrical,
\begin{equation}
I_{ij}=I_{ji},\hspace{1cm}
\mybm{I}_{ij}=-\mybm{I}_{ji}.
\end{equation}
Therefore the diagonal part of the vector current $\mybm{I}_{ii}$ is 
identically zero because of the Fermi statistics.

The four-fermion interactions can be expressed in terms of products of the bilinear currents. Since the Hamiltonian is an SU(2) scalar, only products of scalar-scalar currents or vector-vector ones are allowed. Putting spin indices back into the theory, the momentum-conserving vertices in Eq.~(\ref{realspace1}) can be classified into spin/charge and scalar/vector sectors,
\begin{eqnarray}
{\cal H}^{(1)}_{int} &= & \tilde{c}^{\rho}_{ij} J^{R}_{ij} J^{L}_{ij} 
-\tilde{c}^{\sigma}_{ij} \mybm{J}^{R}_{ij}\cdot  \mybm{J}^{L}_{ij}
\nonumber\\
&+&\tilde{f}^{\rho}_{ij} J^{R}_{ii} J^{L}_{jj} -
\tilde{f}^{\sigma}_{ij} \mybm{J}^{R}_{ii} \cdot \mybm{J}^{L}_{jj}
\nonumber\\
&+& \tilde{s}^{\rho}_{ij} J^{R}_{ij} J^{L}_{\hat{j}\hat{i}} 
-\tilde{s}^{\sigma}_{ij} 
\mybm{J}^{R}_{ij}\cdot  \mybm{J}^{L}_{\hat{j}\hat{i}},
\label{int1}
\end{eqnarray}
where we have used Einstein's convention that summations over the repeated indices are implied. The first two sets of couplings $\tilde{f}_{ij}$ and $\tilde{c}_{ij}$ denote the forward and Cooper scattering amplitudes between bands $i$ and $j$ at generic fillings. However, at half filling $n=1$, due to the particle-hole symmetry near the Fermi surface, additional vertices $\tilde{s}_{ij}$ are also allowed. Note that the vertices $\tilde{s}_{ij}$ involve four different bands and are rather different from the forward and Cooper scattering, $\tilde{f}_{ij}$ and $\tilde{c}_{ij}$, at generic fillings which only involve two bands.

Although the vertices that conserve momentum can be elegantly written down explicitly in terms of SU(2) currents, not all of the couplings are independent. Some of the vertices are doubly counted and should be removed. Besides, various discrete symmetries would further reduce the number of independent couplings. This is actually a good news since we eventually need to solve these couplings numerically. To avoid double counting, one notices that $\tilde{f}_{ii}, \tilde{c}_{ii}$ describe the same vertex. Thus, we choose the diagonal elements of the forward scattering to be zero, i.e. $\tilde{f}_{ii}=0$, to avoid double countings. For the same reason, we choose to set $\tilde{s}_{ii}=0=\tilde{s}_{i\hat{i}}$. Furthermore, under charge conjugation, the indices in the currents switch $J_{ij} \to J_{ji}$, which implies $\tilde{c}_{ij} = \tilde{c}_{ji}$ and 
$\tilde{s}_{ij} = \tilde{s}_{ji}$.  Following similar arguments, parity symmetry (in $x$) implies $\tilde{f}_{ij} =\tilde{f}_{ji}$. Combining the above symmetrical properties with particle-hole symmetry, we found that $A_{ij}=A_{\hat{i}\hat{j}}$ where $A=\{c, f, s \}$. While it is not obvious at this point,  the choice of signs for the scalar and vector couplings in Eq.~(\ref{int1}) is such that they are all positive for the repulsive on-site interaction.

Now we turn to the umklapp interactions which do not conserve momentum exactly (but up to reciprocal lattice vectors). Comparing the difference between interactions in Eqs.~(\ref{realspace1}) and (\ref{realspace2}), we need the other set of SU(2) currents $I_{\mysw{P}ij}$ and $\mybm{I}_{\mysw{P}ij}$ to construct all possible umklapp interactions. Again, the allowed four-fermion interactions can be expressed in terms of the products of SU(2) currents,
\begin{eqnarray}
\mathcal{H}_{int}^{(2)} &=&\frac{\tilde{u}_{ij}^{\rho}}{2}
(I_{\mysw{R}ij}^{\dag} I_{\mysw{L}\hat{i}\hat{j}}
+I_{\mysw{L}\hat{i}\hat{j}}^{\dag} I_{\mysw{R}ij})
-\frac{\tilde{u}_{ij}^{\sigma }}{2}
(\mybm{I}_{\mysw{R}ij}^{\dag} \cdot 
\mybm{I}_{\mysw{L}\hat{i}\hat{j}}
+\mybm{I}_{\mysw{L}\hat{i}\hat{j}}^{\dag }\cdot
\mybm{I}_{\mysw{R}ij})
\nonumber \\
&+&\frac{\tilde{w}_{ij}^{\rho }}{2}
(I_{\mysw{R}i\hat{i}}^{\dag }I_{\mysw{L}j\hat{j}}
+I_{\mysw{L}j\hat{j}}^{\dag }I_{\mysw{R}i\hat{i}})
-\frac{\tilde{w}_{ij}^{\sigma}}{2}
(\mybm{I}_{\mysw{R}i\hat{i}}^{\dag } \cdot \mybm{I}_{\mysw{L}j\hat{j}}
+\mybm{I}_{\mysw{L}j\hat{j}}^{\dag} \cdot 
\mybm{I}_{\mysw{R}i\hat{i}}).
\label{int2}
\end{eqnarray}
Since $\tilde{u}_{ii}, \tilde{w}_{ii}$ describe the same vertex, we choose 
$\tilde{w}_{ii}=0=\tilde{w}_{i\hat{i}}$ to avoid double counting. 
Besides, the product of currents are symmetrical (even though single $\mybm I_{ij}$ is anti-symmetrical), we can choose the couplings to be symmetrical $\tilde{u}_{ij}=\tilde{u}_{ji}$ and $\tilde{w}_{ij}=\tilde{w}_{\hat{i}\hat{j}}$. 
Under parity transformation and charge conjugation, it is easy to show that 
$\tilde{u}_{ij}=\tilde{u}_{\hat{i}\hat{j}}$ and $\tilde{w}_{ij}=\tilde{w}_{ji}$. From Fermi statistics, it is easy to show that $u^{\sigma}_{ii}=0$ and $w^{\sigma}_{mi}=0$, where $m=(N+1)/2$ for OBC's and $m=N/4$ for PBC's (although we haven't discussed the details yet). So far, we only concentrate on the OBC's. In principle, the change of boundary conditions can lead to different sets of allowed interactions in ladder systems. However, as we will explain in the following, it is rather surprising that the above interaction $\mathcal{H} = \mathcal{H}_{int}^{(1)} + \mathcal{H}_{int}^{(2)}$ derived for OBC's is also correct for PBC's.

\subsection{Periodic Boundary Conditions}

For PBCs, the interactions strongly depend on whether the number of 
chains is odd or even. For odd-chain system, half filling is not even 
an umklapp line. As a result, the allowed interactions are greatly simplified and only Cooper and forward scattering in Eq.~(\ref{int1}) are allowed. The detail ground-state phase diagram and RG analysis can be found in the literature\cite{Lin97}. When the number of chains is even, the situation is much more messy. The complication arises from more identities at half filling,
\begin{eqnarray}
k_{Fi}-k_{F\pm i}&=&0,
\label{identity3}\\
k_{Fi}+k_{F\pm \hat{i}}&=&\pi,
\label{identity4}
\end{eqnarray}
where $\hat{i} \equiv {\rm sign}(i)(N/2) -i$. The $\pm$ sign in the 
identities and the dependence of sign$(i)$ in the definition of 
$\hat{i}$ makes the analysis a bit messy. Note that these additional identities reflect the peculiar properties of the nested Fermi surface at half filling. 

As in previous subsection, we classify the vertices by $(n_{x}, n_{y})$, where $n_{x},n_{y}$ can take values, $0, \pm 1, \pm 2$. However, $n_{x}=\pm 2$ is
ruled out because it corresponds to the interactions between completely
filled bands which, as mentioned before, do not participate the 
low-energy physics here. For $n_{y}=\pm 2$, all the bands are lying 
on the boundary of Brillouin zone, i.e. $i_{a}= \pm N/2$, which actually 
indicates the same band and thus included in the $n_{y}=0$ case. Therefore, only four kinds of vertices are allowed: (0,0), $(0, \pm 1)$, $(\pm 1,0)$, and 
$(\pm 1, \pm 1)$.

For $(n_{x}, n_{y})=(0, 0)$, the momenta is conserved in both 
directions. We can construct the linear combination from Eq.~(\ref{identity3})
\begin{equation}
k_{Fi}-k_{F\pm i}+k_{Fj}-k_{F\pm j}=0,
\end{equation}
which gives rise to two sets of solutions. The first set of the solutions is,
\begin{eqnarray}
  (P_1, P_2)&=&(P_3, P_4)
  \nonumber\\
  (i_1,i_2) &=&(\pm i_3, \pm i_4),
 \end{eqnarray}
and the second set is,
\begin{eqnarray}
  P_1=\overline{P}_2, &\qquad& P_3=\overline{P}_4;
  \nonumber\\
  i_1= \pm i_2, &\qquad& i_3= \pm i_4.
\end{eqnarray}
However, since the transverse momenta is also conserved, the solutions
need to satisfy Eq.~(\ref{Ky_cons}) as well, which fixes the ambiguity of 
the signs. The resulted vertices are again the Cooper and forward scattering.

Following the same spirit, other interactions can be constructed by different linear combinations of the identities arisen from the nested Fermi surface. The linear combination from Eq.~(\ref{identity4}) gives
\begin{equation}
k_{Fi}+k_{F\pm \hat{i}}-k_{Fj}-k_{F\pm \hat{j}}=0.
\end{equation}
The conservation of transverse momenta puts a more severe 
constraint in this case. For forward and Cooper vertices, it rules out the annoying sign ambiguity and leads to the same vertices as in OBC's. Here, 
not only pin down the signs, it also puts constraints on the band indices
\begin{eqnarray}
  P_1=\overline{P}_2, &&P_3=\overline{P}_4,
  \nonumber\\
 (P_1, P_2)&=&(P_3, P_4),
 \nonumber\\ 
 (i_1,i_2) &=&(\hat{i}_{4}, \hat{i}_{3}), \hspace{1cm}
 i_{1} \cdot i_{2}>0,
\end{eqnarray}
where $i_{1}, i_{2}$ must have the {\em same} sign! This strange constraint is quite tough to handle when one wants to perform RG analysis. But, as will become evident in a moment, The other missing vertices with opposite signs $i_{1} \cdot i_{2} <0$ appears in another sector and the strange constraint is indeed an artifact of how we classify the vertices.

It is quite similar to find the $(0, \pm 1)$ vertices. In stead of repeating the whole calculations again,  we briefly sketch the steps to obtain the solutions. First, the conservation of $k_{x}$ momenta is the same as for (0,0) vertices. The difference arises because the right-hand side of Eq.~(\ref{Ky_cons}) is not zero. After a liitle algebra, three sets of solutions are found. The first set of solutions is
\begin{eqnarray}
  (P_1, P_2) &=& (P_3, P_4),
  \nonumber\\
  (i_1,i_2) &=& (\bar{i_3}, \bar{i_4}), \hspace{1cm}
  |i_{1}-i_{2}|=\frac{N}{2};
 \end{eqnarray}
and the second one is,
\begin{eqnarray}
  P_1=\overline{P}_2, &\quad & P_3=\overline{P}_4;
  \nonumber\\
  i_1= \bar{i_2}, &\quad & i_3= \bar{i_4}, \hspace{1cm}
  |i_{1}-i_{3}|=\frac{N}{2}.
\end{eqnarray}
The last one is
\begin{eqnarray}
  P_1=\overline{P}_2, &&P_3=\overline{P}_4;
  \nonumber\\
 (P_1, P_2)&=&(P_3, P_4);
 \nonumber\\ 
 (i_1,i_2) &=&(\hat{i}_{4}, \hat{i}_{3}), \hspace{1cm}
 i_{1} \cdot i_{2}<0,
\end{eqnarray}
where $i_{1}, i_{2}$ must have the {\em different} signs. One notices 
the last set of solution has appeared in the (0,0) vertex except the 
constraint on the band indices is different. 

It is straightforward to show that the first two sets of solutions are 
actually included in the last one already. With the simplification, we 
can write down the Hamiltonian density for $(0,0)$ and 
$(0, \pm1)$ vertices together in terms of products of SU(2) currents. Now it is clear that these vertices are described by ${\cal H}^{(1)}_{int}$ in Eq.~(\ref{int1}) -- the same as in OBC's!! Similarly, when combining the $(\pm1, 0)$ and $(\pm 1, \pm 1)$ vertices, we end up with the Hamiltonian ${\cal H}^{(2)}_{int}$ in Eq.~(\ref{int2}). It is worth commenting that the extra subtleties coming from the conservation of transverse momentum make the search for all allowed vertices much more difficult and messy. But, at the end, the effective Hamiltonian takes exactly the same form as in OBC's, except the definition of dual points on the Fermi surface are different. At the time of writing, we do not fully understand the reason behind this coincident. 

\section{One-Loop RG}

In stead of following the conventional diagrammatic technique, we derive the one-loop RG equations by operator product expansions (OPE) of the bilinear currents. Since the leading divergence is logarithmic, the one-loop RG equations are universal despite of different schemes. In fact, the universality is crucial since we later show that the exponents of anomalous scaling are directly determined by the coefficients in the RG equations. Since the exponents of various power laws are robust, we expect the coefficients of the one-loop RG equations to be universal, which agrees with the logarithmic divergences of marginal interaction (four-fermion vertices) in Q1D systems.

In the following, we outline the steps to obtain the RG equations and also derive one of the renormalized coupling as a simple example to familiarize readers with the technique. To proceed, it is convenient to introduce the partition function in path integral formalism,
\begin{eqnarray}
Z\; &=&\;\int D[\overline{\psi}\psi] e^{-S_{0}-S_{int}}
\nonumber \\
&=&\int\!D[\overline{\psi}\psi] e^{-S_{0}}
\frac{\int\!D[\overline{\psi}\psi] e^{-S_{0}} e^{-S_{int}}}{\int\!D[\overline{\psi}\psi] e^{-S_{0}}}=\;Z_{0}\;\langle
e^{-S_{int}}\rangle _{0}\;,
\end{eqnarray}
where $Z_{0}$ is the partition function without interactions and $\langle
e^{-S_{int}}\rangle _{0}$ with the subscript 0 denotes the average over the
unperturbed action $S_{0}$ only. We can reexponentiated it by the cumulant
expansion,
\begin{equation}
\langle e^{-S_{int}}\rangle _{0}\;=\;\exp \left\{ \langle -S_{int}\rangle
_{0}+\frac{1}{2}[\langle S_{int}^{2}\rangle _{0}-\langle S_{int}\rangle
_{0}^{2}]\ +\ O(S_{int}^{3})\right\} \;,
\end{equation}
The only nontrivial term is $\langle S_{int}^{2}\rangle _{0}$ which will
renormalize the other interactions. To one-loop order,
\begin{equation}
\langle -S_{int}^{2}\rangle _{0}\;=\;\int_{a}^{\infty }\prod_{i=1,2}\;d\tau
_{i}\;dx_{i}\;\langle \;\mathcal{H}_{int}(\tau _{1},x_{1})\;
\mathcal{H}_{int}(\tau _{2},x_{2})\;\rangle _{0}\;,
\end{equation}
the notation $\int_{a}^{\infty }\equiv \ \int_{A<\mid x_{1}-x_{2}\mid <B}$. Each integral has two parts: long-wavelength modes, $ba<\mid
x_{1}-x_{2}\mid <\infty $, and short-wavelength modes, $a<\mid
x_{1}-x_{2}\mid <ba$, where $b>1$ is the rescaling parameter. When we
integrate out those short-wavelength modes, since all fields are at nearby
space points, we can use the OPE to replace the product of them by a series of local operators, that is,
\begin{eqnarray}
\int_{a}^{ba}\prod_{i=1,2}\;d\tau _{i}\;dx_{i}\;\langle \;
\mathcal{H}_{int}(\tau _{1},x_{1})\;
\mathcal{H}_{int}(\tau _{2},x_{2})\;\rangle _{0}
\simeq \;\int d\tau \;dx\;\langle \delta \mathcal{H}_{int}\rangle _{0}\;.
\end{eqnarray}
The effective interaction $\delta \mathcal{H}_{int}$ has the same form as
the original Hamiltonian and renormalize corresponding couplings when the short-range fluctuations are progressively integrated out. Studying how the couplings get renormalized, the one-loop RG equations can be derived.

Here we demonstrate how to calculate the effective Hamiltonian $\delta
\mathcal{H}_{int}$ with the use of OPE's explicitly. First of all, when chiral fermion fields $\psi_{R/Li\alpha }$ (with the same chirality) are brought closer together in space and time, the leading singularities are simple poles,
\begin{eqnarray}
\psi _{Ri\alpha }(x,\tau ) \psi _{Rj\beta }^{\dag }(0,0)
&\sim &\frac{\delta _{ij}\delta _{\alpha \beta }}{z_{i}} + O(1),
\\
\psi _{Li\alpha }(x,\tau )\psi _{Lj\beta }^{\dag }(0,0)
&\sim & \frac{\delta _{ij}\delta _{\alpha \beta }}{z_{i}^{*}}+ O(1),
\end{eqnarray}
where $z_{i}=2\pi \ (v_{i}\tau-ix)$. The above equations are valid as long as $(x,\tau )$ close enough to $(0,0)$. The full set of OPE can be obtained, making use of the above identities. For instance, when two right-moving currents $J_{Rij}$ and $J_{Rkl}$ are brought closer in space and time, their product can be approximated by another operator,
\begin{eqnarray}
J_{ij}(x,\tau ) J_{kl}(0,0) &= & \psi _{i\alpha }^{\dag }(x,\tau )\psi
_{j\alpha }(x,\tau )\ \psi _{k\beta }^{\dag }(0,0)\psi _{l\beta }(0,0)
\nonumber\\
&& \hspace{-2.2cm}
\sim \left\langle \psi _{i\alpha }^{\dag }(z_{i})
\psi _{l\beta}(0)\right\rangle
\psi _{j\alpha }(z_{j}) \psi _{k\beta }^{\dag}(0)
+\left\langle \psi _{j\alpha }(z_{j})
\psi _{k\beta }^{\dag}(0)\right\rangle
\psi _{i\alpha }^{\dag }(z_{i})\psi _{l\beta}(0)
\nonumber \\
&& \hspace{-2.2cm} \sim- \frac{\delta _{il}\delta _{\alpha \beta }}{z_{i}}
\psi_{k\beta }^{\dag }(0) \psi _{j\alpha }(z_{j})
+\frac{\delta _{jk}\delta _{\alpha \beta }}{z_{j}}
\psi _{i\alpha}^{\dag }(z_{i})\psi _{l\beta }(0)
= -\;\frac{\delta _{il}}{z_{i}}J_{kj}+\frac{\delta _{jk}}{z_{j}}J_{il}.
\end{eqnarray}
Here, for notational convenience, the subscript of the chirality is suppressed.
The others operator products can be calculated in the similar way. The full set of OPE can be found in Appendix A.

With these OPE's, we now perform explicit calculations for a typical term
in $\langle -S_{int}^{2}\rangle _{0}$, say
\begin{equation}
\frac{1}{8}\;\tilde{c}_{ij}^{\rho }\;\tilde{c}_{kl}^{\rho }\;\int_{z,w}\
\langle
\;J_{ij}^{R}(z)\;J_{ij}^{L}(z)\;J_{kl}^{R}(w)\;J_{kl}^{L}(w)\;\rangle
\end{equation}
where $\int_{z,w}$ denotes a four-dimensional integral over two complex
planes $z$ and $w$. In the RG procedure, we integrate out degrees of freedom at short-length scale, i.e. where the complex varibles $z$ and $w$ are close together. Replacing the operator products by equivalent local operators, the OPE's give  
\begin{eqnarray*}
&&\frac{1}{8}\;\tilde{c}_{ij}^{\rho }\;\tilde{c}_{kl}^{\rho
}\;\int_{z,w} \left\langle 
\left[ \frac{\delta _{jk}}{(z_{j}-w_{j})}J_{il}^{R}
-\frac{\delta _{il}}{(z_{i}-w_{i})}J_{kj}^{R}\right]
\left[ \frac{\delta _{jk}}{(z_{j}^{*}-w_{j}^{*})}J_{il}^{L}
-\frac{\delta _{il}}{(z_{i}^{*}-w_{i}^{*})}J_{kj}^{L}\right] \right\rangle  
\nonumber\\
&&
= \int_{z,w} \frac{1}{8} \tilde{c}_{ij}^{\rho} \tilde{c}_{jl}^{\rho}\:
\frac{1}{\mid z_{j}-w_{j}\mid ^{2}} J_{il}^{R}(w)J_{il}^{L}(w)
+\frac{1}{8} \tilde{c}_{ij}^{\rho} \tilde{c}_{ki}^{\rho}\:
\frac{1}{\mid z_{i}-w_{i}\mid ^{2}}
J_{kj}^{R}(w) J_{kj}^{L}(w)
\nonumber \\
&& 
-\frac{1}{8} \tilde{c}_{ij}^{\rho}\tilde{c}_{ji}^{\rho} \:
\left[ \frac{1}{(z_{i}-w_{i})(z_{j}^{*}-w_{j}^{*})}
J_{jj}^{R}(w)\;J_{ii}^{L}(w)+
\frac{1}{(z_{j}-w_{j})(z_{i}^{*}-w_{i}^{*})}
J_{ii}^{R}(w)\;J_{jj}^{L}(w) \right].
\end{eqnarray*}

The integration over the fast modes is elementary,
\begin{equation}
\int_{a<\mid x\mid <ba}\!\! dx \int_{-\infty }^{\infty }d\tau\:
\frac{1}{z_{i}z_{j}^{*}} = \frac{\ln \;b}{\pi (v_{i}+v_{j})}
= \frac{dl}{\pi(v_{i}+v_{j})}.
\end{equation}
After integrating over the short-range fluctuations and collecting terms together, the renormalized action is,
\begin{equation}
\sum_{k}\frac{1}{4}\;\tilde{c}_{ik}^{\rho
}\;\tilde{c}_{kj}^{\rho }\;\frac{dl}{2\pi v_{k}}\;\int_{z}\;J_{ij}^{R}\
J_{ij}^{L}-\;\frac{1}{4}\ (\tilde{c}_{ij}^{\rho })^{2}\;\frac{dl}{\pi
(v_{i}+v_{j})}\;\int_{z}J_{ii}^{R}\ J_{jj}^{L}.
\end{equation}
Compared with the bare Hamiltonian, it contributes to renormalization of Cooper scattering $c^{\rho}_{ij}$ and forward scattering $f^{\rho}_{ij}$,
\begin{eqnarray}
-\ \delta \tilde{c}_{ij}^{\rho }\; &=&\frac{1}{4}\;\sum_{k}\;
\tilde{c}_{ik}^{\rho }\;\tilde{c}_{kj}^{\rho }\;\frac{dl}{2\pi v_{k}},
\\
-\;\delta {\tilde{f}}_{ij}^{\rho }\; &=&-\;\frac{1}{4}\ 
(\tilde{c}_{ij}^{\rho })^{2}\frac{dl}{\pi (v_{i}+v_{j})}.
\end{eqnarray}
To simplify the notation, we define $A_{ij}=\tilde{A}_{ij}/\pi (v_{i}+v_{j})$, where $A=\{b,f,s,u,w\}$ and $\alpha _{ij,k}\equiv
(v_{i}+v_{k})(v_{j}+v_{k})/2v_{k}(v_{i}+v_{j})$. The RG equations are,
\begin{eqnarray}
\dot{c}_{ij}^{\rho}\ &=&-\sum_{k}\frac14 \alpha _{ij,k}
c_{ik}^{\rho} c_{kj}^{\rho} + ... \:\:,
\label{rg1} \\
\dot{f}_{ij}^{\rho} &=&\frac{1}{4} (c_{ij}^{\rho })^{2} + ... \:\:,
\label{rg2}
\end{eqnarray}
where $\dot{A} \equiv dA/dl$. One can follow similar steps to compute the renormalization of all couplings to one-loop order. After lengthy algebra, the whole set of coupled RG equations is derived and presented in Appendix B.

Note that the renormalization of interacting vertices is described by the coupled nonlinear differential equations such as Eqs.~(\ref{rg1}) and (\ref{rg2}). From mathematical viewpoint, the flows of these nonlinear equations should exhibit chaotic behavior. However, numerical studies of these messy equations give rather simple and robust phase diagrams. Somehow, the flows is ``protected'' from the generic chaotic fate. It turns out that there exists hidden structure behind the messy-looking RG equations. After appropriate rescaling, the whole set of equations can be derived from a single potential which we nickname as RG potential. As the old saying states -- seeing is believing, we will present two examples of RG potential in the next section first. Then, in following up section, we give the general proof for the existence of RG potential but construct it explicitly.

\section{Hidden Structure in RG Flows}

The first example we plan to present is the two-leg ladder at half filling. After chiral field decomposition, the kinetic energy consists of four pairs of Dirac fermions,
\begin{equation}
{\cal H}_0= \psi^{\dag}_{\mysw{R}i\alpha} (-iv_i \partial_x) \psi^{}_{\mysw{R}i\alpha} + \psi^{\dag}_{\mysw{L}i\alpha} (iv_i \partial_x) \psi^{}_{\mysw{L}i\alpha},
\end{equation}
where $i=1,2$ denotes the band index and $\alpha = \uparrow, \downarrow$ are the spin indices. Due to the symmetry between bonding and antibonding bands, the Fermi velocities are equal $v_{1}=v_{2}=v$. Since there are only two interacting bands, the interactions are not as complicated as we derived before,
\begin{eqnarray}
{\cal H}_{int} &=& \tilde{c}^{\rho}_{ij} J_{\mysw{R}ij} J_{\mysw{L}ij} - \tilde{c}^{\sigma}_{ij} \mybm{J}_{\mysw{R}ij} \cdot \mybm{J}_{\mysw{L}ij}+ \tilde{f}^{\rho}_{ij} J_{\mysw{R}ii} J_{\mysw{L}jj} - \tilde{f}^{\sigma}_{ij} \mybm{J}_{\mysw{R}ii} \cdot \mybm{J}_{\mysw{L}jj}
\nonumber\\
&+& \frac{\tilde{u}^{\rho}_{ij}}{2} (I_{\mysw{R}ij} I^{\dag}_{\mysw{L}ji} + I_{\mysw{L}ij} I^{\dag}_{\mysw{R}ji}) - \frac{\tilde{u}^{\sigma}_{ij}}{2} (\mybm{I}_{\mysw{R}ij} \cdot \mybm{I}^{\dag}_{\mysw{L}ji} + \mybm{I}_{\mysw{L}ij} \cdot \mybm{I}^{\dag}_{\mysw{R}ji}).
\end{eqnarray}
Summation over the band indices $i,j = 1,2$ are implied. Note that the other two kinds of vertices, $s^{\rho,\sigma}_{ij}$ and $w^{\rho,\sigma}_{ij}$, do not show up because of the number of interacting bands is only two here. 

Since the Fermi velocities in bonding and antibonding bands are the same, we can choose the Fermi velocity $2\pi v =1$ to simplify the numerical factors. To give the readers a flavor of the complexity of the coupled non-linear RG equations, we write them down explicitly here,
\begin{eqnarray}
\frac{d\tilde{c}^{\rho}_{11}}{dl} &=& - \frac14 (\tilde{c}^{\rho}_{12})^{2}
-\frac{3}{4}(\tilde{c}^{\sigma}_{12})^{2}
+ \frac14 (\tilde{u}^{\rho}_{12})^{2}+\frac{3}{4}(\tilde{u}^{\sigma}_{12})^{2},
\\
\frac{d\tilde{c}^{\sigma}_{11}}{dl}&=& -(\tilde{c}^{\sigma}_{11})^{2}
-\frac12 \tilde{c}^{\rho}_{12}\tilde{c}^{\sigma}_{12}-\frac12 (\tilde{c}^{\sigma}_{12})^{2}
-\frac12 \tilde{u}^{\rho}_{12}\tilde{u}^{\sigma}_{12}
-\frac12 (\tilde{u}^{\sigma}_{12})^{2},
\\ 
\frac{d\tilde{c}^{\rho}_{12}}{dl}&=& -\frac12 \tilde{c}^{\rho}_{11}\tilde{c}^{\rho}_{12}
-\frac{3}{2} \tilde{c}^{\sigma}_{11}\tilde{c}^{\sigma}_{12}
+\frac12 \tilde{c}^{\rho}_{12}\tilde{f}^{\rho}_{12}
+\frac{3}{2} \tilde{c}^{\sigma}_{12}\tilde{f}^{\sigma}_{12}
+\tilde{u}^{\rho}_{12}\tilde{u}^{\rho}_{11},
\\
&&\hspace{-2cm}
\frac{d\tilde{c}^{\sigma}_{12}}{dl} = -\frac12 \tilde{c}^{\rho}_{11}\tilde{c}^{\sigma}_{12}
-\frac12 \tilde{c}^{\rho}_{12}\tilde{c}^{\sigma}_{11}
-\tilde{c}^{\sigma}_{12}\tilde{c}^{\sigma}_{11}
+ \frac12 \tilde{f}^{\rho}_{12}\tilde{c}^{\sigma}_{12}
+\frac12 \tilde{c}^{\rho}_{12}\tilde{f}^{\sigma}_{12}
-\tilde{c}^{\sigma}_{12}\tilde{f}^{\sigma}_{12}
+\tilde{u}^{\rho}_{11}\tilde{u}^{\sigma}_{12},
\\ 
\frac{d\tilde{f}^{\rho}_{12}}{dl}&=& \frac14 (\tilde{c}^{\rho}_{12})^{2}
+\frac{3}{4}(\tilde{c}^{\sigma}_{12})^{2}
+\frac14 (\tilde{u}^{\rho}_{12})^{2}
+\frac{3}{4}(\tilde{u}^{\sigma}_{12})^{2}
+(\tilde{u}^{\rho}_{11})^{2},
\\
\frac{d\tilde{f}^{\sigma}_{12}}{dl}&=& -(\tilde{f}^{\sigma}_{12})^{2}
+\frac12 \tilde{c}^{\rho}_{12}\tilde{c}^{\sigma}_{12}
-\frac12 (\tilde{c}^{\sigma}_{12})^{2} 
+ \frac12 \tilde{u}^{\rho}_{12}\tilde{u}^{\sigma}_{12}
-\frac12 (\tilde{u}^{\sigma}_{12})^{2},
\\
\frac{d\tilde{u}^{\rho}_{11}}{dl} &=& \tilde{f}^{\rho}_{12}\tilde{u}^{\rho}_{11}
+\frac12 \tilde{c}^{\rho}_{12}\tilde{u}^{\rho}_{12}
+\frac{3}{2} \tilde{c}^{\sigma}_{12}\tilde{u}^{\sigma}_{12},
\\
\frac{d\tilde{u}^{\rho}_{12}}{dl} &=& \tilde{c}^{\rho}_{12}\tilde{u}^{\rho}_{11}
+\frac12 \tilde{c}^{\rho}_{11}\tilde{u}^{\rho}_{12}
-\frac{3}{2}\tilde{c}^{\sigma}_{11}\tilde{u}^{\sigma}_{12}
+\frac12 \tilde{f}^{\rho}_{12}\tilde{u}^{\rho}_{12}
+\frac{3}{2}\tilde{f}^{\sigma}_{12}\tilde{u}^{\sigma}_{12},
\\
&&\hspace{-2cm}
\frac{d\tilde{u}^{\sigma}_{12}}{dl} = \frac12 \tilde{c}^{\rho}_{11}\tilde{u}^{\sigma}_{12}
-\frac12 \tilde{c}^{\sigma}_{11}\tilde{u}^{\rho}_{12}
-\tilde{c}^{\sigma}_{11}\tilde{u}^{\sigma}_{12}
+\frac12 \tilde{f}^{\sigma}_{12}\tilde{u}^{\rho}_{12}
+\frac12 \tilde{f}^{\rho}_{12}\tilde{u}^{\sigma}_{12}
-\tilde{f}^{\sigma}_{12}\tilde{u}^{\sigma}_{12}
+\tilde{c}^{\sigma}_{12}\tilde{u}^{\rho}_{11}.
\end{eqnarray}
Neither do the above equation look inspiring, nor do they seem to reveal any structure/symmetry behind. However, if we rescale the couplings in the following way,
\begin{eqnarray}
&& \bigg[ \tilde{u}^{\rho}_{11}, \tilde{u}^{\rho}_{12},
\tilde{c}^{\rho}_{11}, \tilde{c}^{\rho}_{12},
\tilde{f}^{\rho}_{12}, \tilde{u}^{\sigma}_{12},
\tilde{c}^{\sigma}_{11}, \tilde{c}^{\sigma}_{12},
\tilde{f}^{\sigma}_{12} \bigg] \qquad \mbox{(original $g_{i}$)}
\nonumber\\
=\: && \bigg[ \frac{u^{\rho}_{11}}{\sqrt{2}}, u^{\rho}_{12},
c^{\rho}_{11}, c^{\rho}_{12},
f^{\rho}_{12}, \frac{u^{\sigma}_{12}}{\sqrt{3}},
\frac{c^{\sigma}_{11}}{\sqrt{3}}, \frac{c^{\sigma}_{12}}{\sqrt{3}},
\frac{f^{\sigma}_{12}}{\sqrt{3}} \bigg] \qquad \mbox{(rescaled $h_{i}$)}.
\end{eqnarray}
In terms of the rescaled couplings $h_{i}$, the coupled non-linear RG equations can be mapped into potential flows,
\begin{eqnarray}
\frac{dh_{i}}{dl} = - \frac{\partial V}{\partial h_{i}},
\end{eqnarray}
where all the RG flows are captured by a single potential,
\begin{eqnarray}
V(h_i) &=& -\frac{1}{3\sqrt{3}} (c^{\sigma}_{11})^3 - \frac{1}{3\sqrt{3}} (f^{\sigma}_{12})^3
\nonumber\\
&+& \frac14 (f^{\rho}_{12} - c^{\rho}_{11}) \left[ (c^{\rho}_{12})^2 + (c^{\sigma}_{12})^2\right]
+ \frac14 (f^{\rho}_{12} + c^{\rho}_{11}) \left[ (u^{\rho}_{12})^2 + (u^{\sigma}_{12})^2\right]
\nonumber\\
&+& \frac12 (f^{\sigma}_{12} - c^{\sigma}_{11}) \left[ c^{\rho}_{12} c^{\sigma}_{12} + u^{\rho}_{12} u^{\sigma}_{12} \right]
- \frac{1}{2\sqrt{3}} (f^{\sigma}_{12} + c^{\sigma}_{11}) \left[ (c^{\sigma}_{12})^2 + (u^{\sigma}_{12})^2\right]
\nonumber\\
&+& \frac{1}{\sqrt{2}} u^{\rho}_{11} u^{\sigma}_{12} c^{\sigma}_{12} + \frac{1}{\sqrt{2}} u^{\rho}_{11} u^{\rho}_{12} c^{\rho}_{12} + \frac12 (u^{\rho}_{11})^2 f^{\rho}_{12}.
\label{TwoLegRGP}
\end{eqnarray}
This is truly remarkable that the complexity of the RG equations can be removed by simple rescaling. Since the flows under the influence of potential can be viewed as the trajectories of an overdamped particle in multi-dimensional coupling space, we know the ultimate fate of the RG flows is nothing but chasing the minima of the potential $V$. To one-loop order, the potential does not have any minimum (except the trivial one at $h_{i}=0$). Therefore, the flows are attracted to the fastest descending asymptotes that eventually go into the strong coupling regime. It is rather easy to find out the asymptotes where the couplings keep constant ratios. That's why chaos are never spotted in numerical studies for the RG equations of Q1D systems.

One may question that the two-leg ladder at half filling may be a special case and the generic Q1D systems are not necessarily described by the RG potential. Therefore, it is insightful to consider the doped $N$-chain ladder at generic fillings. The kinetic energy is described by $2N$ (the factor of 2 comes from spin) pairs of Dirac fermions with different velocities $v_i$. At generic fillings, the shape of Fermi surface does not have any special symmetry. As a result, only forward and Cooper vertices are allowed,
\begin{eqnarray}
{\cal H}_{int} &=& \tilde{c}^{\rho}_{ij} J_{\mysw{R}ij} J_{\mysw{L}ij} 
- \tilde{c}^{\sigma}_{ij} \mybm{J}_{\mysw{R}ij} \cdot \mybm{J}_{\mysw{L}ij}
+ \tilde{f}^{\rho}_{ij} J_{\mysw{R}ii} J_{\mysw{L}jj} 
- \tilde{f}^{\sigma}_{ij} \mybm{J}_{\mysw{R}ii} \cdot \mybm{J}_{\mysw{L}jj},
\end{eqnarray}
where summations over the band indices $i,j = 1,2,...,N$ are implicitly indicated. To avoid repetition, we do not intend to write down the RG equations and redirect the interested readers to Appendix B. The RG equations for the doped Q1D ladder can be obtained by setting the other couplings $s^{\rho,\sigma}_{ij} = u^{\rho,\sigma}_{ij} =w^{\rho,\sigma}_{ij}=0$ in the equations. 

The presence of different Fermi velocities makes the search for appropriate rescaling factors not a trivial task,
\begin{eqnarray}
\tilde{c}^{\rho}_{ii} = 2\pi \left( 4\sqrt{2} v_{i} \right) c^{\rho}_{ii},
&\qquad&
\tilde{c}^{\sigma}_{ii} = 2 \pi \left( \frac{4\sqrt{2}}{\sqrt{3}} v_{i} \right) c^{\sigma}_{ii},
\nonumber\\
\tilde{c}^{\rho}_{ij} = 2\pi \left( 4\sqrt{v_i v_j}\right) c^{\rho}_{ij},
&\qquad&
\tilde{c}^{\sigma}_{ij} = 2\pi \left( \frac{4}{\sqrt{3}} \sqrt{v_i v_j} \right) c^{\sigma}_{ij},
\nonumber\\
\tilde{f}^{\rho}_{ij} = 2\pi \left( 4\sqrt{v_i v_j}\right) f^{\rho}_{ij},
&\qquad&
\tilde{f}^{\sigma}_{ij} = 2\pi \left( \frac{4}{\sqrt{3}} \sqrt{v_i v_j} \right) f^{\sigma}_{ij}.
\end{eqnarray}
It is rather remarkable that, despite of the presence of difference Fermi velocities, the RG potential still exists! The whole set of complicated RG equations can be derived from the potential below
\begin{eqnarray}
V(g_i) &=& -\sum_i \frac{4\sqrt{2}}{3\sqrt{3}} (c^{\sigma}_{ii})^3 
- \sum_{i<j} \frac{4}{3\sqrt{3}} \gamma_{ij} (f^{\sigma}_{ij})^3
\nonumber\\
&+& \sum_{i<j} (c^{\rho}_{ij})^2 \left[ -\frac{1}{\sqrt{2}} c^{\rho}_{ii}
-\frac{1}{\sqrt{2}} c^{\rho}_{jj} + \gamma_{ij} f^{\rho}_{ij} \right]
\nonumber\\
&+& \sum_{i<j} (c^{\sigma}_{ij})^2 \left[ -\frac{1}{\sqrt{2}} c^{\rho}_{ii}
-\frac{1}{\sqrt{2}} c^{\rho}_{jj} + \gamma_{ij} f^{\rho}_{ij} \right]
\nonumber\\
&+& \sum_{i<j} (c^{\sigma}_{ij})^2 \left[ -\frac{\sqrt{2}}{\sqrt{3}} c^{\sigma}_{ii}
-\frac{\sqrt{2}}{\sqrt{3}} c^{\sigma}_{jj} - \frac{2}{\sqrt{3}} \gamma_{ij} f^{\sigma}_{ij} \right]
\nonumber\\
&+& \sum_{i<j} c^{\rho}_{ij}c^{\sigma}_{ij} \left[ -\sqrt{2} c^{\sigma}_{ii}
-\sqrt{2} c^{\sigma}_{jj} + 2 \gamma_{ij} f^{\sigma}_{ij} \right]
\nonumber\\
&-& \sum_{i<j<k} c^{\rho}_{ij} c^{\rho}_{jk} c^{\rho}_{ik}
- \frac{2}{\sqrt{3}} \sum_{i<j<k} c^{\sigma}_{ij} c^{\sigma}_{jk} c^{\sigma}_{ik}
\nonumber\\
&-&  \sum_{i<j<k} [c^{\rho}_{ij} c^{\sigma}_{jk} c^{\sigma}_{ik}+c^{\sigma}_{ij} c^{\rho}_{jk} c^{\sigma}_{ik}+c^{\sigma}_{ij} c^{\sigma}_{jk} c^{\rho}_{ik}],
\label{NLegRGP}
\end{eqnarray}
where the Fermi velocities enter the RG potential only through the ratio of geometric and algebraic averages for different pairs of Fermi velocities,
\begin{equation}
\gamma_{ij} = \frac{2\sqrt{v_i v_j}}{v_i +v_j}.
\end{equation}
Since both the half-filled two-leg ladder and the doped $N$-chain system at generic fillings are governed by the RG potential, it motivated us to look for a more general way to construct the potential in next section.

\section{RG Potential}

To prove the existence of the RG potential, it is helpful to study the general feature of one-loop RG equations of the Q1D systems first. In weak
coupling, the most relevant interactions are the marginal four-fermion interactions, described by a set of dimensionless couplings $g_i$.
The RG transformation to the one-loop order is described by a set of coupled non-linear first-order differential equations
\begin{equation}
\frac{dg_i}{dl} = M^{jk}_{i} g_j g_k \equiv F_i,
\label{OneLoop}
\end{equation}
where the coefficients $M^{jk}_{i} = M^{kj}_{i}$ are symmetrical
by construction. These constant tensors $M^{jk}_{i}$ completely
determine the RG flows.

The solution for Eq.~(\ref{OneLoop}) can be viewed as
the trajectory of a strongly overdamped particle under the
influence of the external force $F_i$ in the multi-dimensional
coupling space. Naively, one might rush to the conclusion that the
conditions for the existence of a potential requires are,
\begin{equation}
\frac{\partial F_i}{\partial g_j} - \frac{\partial F_j}{\partial g_i} =0, \hspace{4mm}
\rightarrow \hspace{4mm} M_{i}^{jk} = M_{j}^{ik},
\end{equation}
which implies that the tensor $M_{i}^{jk}$is totally symmetric. It
is straightforward to check that the RG equations for the Q1D
systems {\em do not} satisfy this criterion \cite{Lin97,Ledermann00}.

However, under general linear transformations of the couplings $h_i(l) = L_{ij} g_{j}(l)$, the coefficients $M_{i}^{jk}$ transform into a new set of coefficients $N_{i}^{jk}$, which may become symmetric. For convenience, we introduce a set of matrices $[\bm{M}(k)]_{ij}\equiv M_{i}^{jk}$ to represent the coefficients. The symmetric criterion for $N_{i}^{jk}=N_{j}^{ik}$ requires the existence of a constant matrix $\bm{L}$ which satisfies the following constraints (for all $k$!), 
\begin{equation}
\bm{M}(k)^T = (\bm{L}^{T} \bm{L}) \bm{M}(k)( \bm{L}^{T} \bm{L})^{-1},
\end{equation}
where superscript $T$ means transpose. In general, there is no guarantee why the strongly over-determined constraints would allow a solution for $\bm{L}$. In fact, it is a nontrivial task to just prove/disprove whether the desired linear transformation $\bm{L}$ exists. Surprisingly, for the Q1D systems, the hunt for the solution greatly simplifies if we formulate the problem in terms of Majorana fermions. The desired transformation becomes diagonal, $L_{ij} = r_{i} \delta_{ij}$, where $r_i$ is a set of rescaling factors and leads to the totally symmetric coefficients,
\begin{equation}
N^{jk}_{i} = \left(\frac{r_i}{r_j r_k}\right) M^{jk}_{i}.
\end{equation}
So the search for the potential is now nailed down to find a set of rescaling factors $r_i$ in the Majorana representation. In later section, we demonstrate how to construct the RG potential explicitly in the Majorana representation. In fact, one can also construct the potential $V(h_i)$ directly from the RG equations for doped\cite{Lin97} and half-filled\cite{Ledermann00} Q1D systems. Both approaches lead the the same result and the detail work will be described elsewhere.

It is important to discuss a special set of analytic solution of Eq.~(\ref{OneLoop}), which is closely related to the scaling Ansatz in Eq.~(\ref{Scaling}). Suppose the initial values of the couplings are $ g_{i}(0) =  G_{i} g(0) $, where $g(0)=U \ll 1$ and $G_{i}$ are order-one constants satisfying the non-linear algebraic constraint,
\begin{equation}
G_{i} = M^{jk}_{i} G_{j} G_{k}.
\label{Algebraic}	
\end{equation}
It is straightforward to show that the ratios between couplings 
remain the same and the complicated equations reduce to single 
one,
\begin{eqnarray}
\frac{dg}{dl} = g^{2}.
\end{eqnarray}
For repulsive interaction $U>0$, the above equation can be solved easily $g(l) = 1/(l_{d}-l)$, where the divergent length scale $l_{d} = 1/U$. Note that this implies the ratios of different couplings remain fixed in the RG flows,
\begin{eqnarray}
g_{i}(l) = \frac{G_{i}}{l_{d}-l}.
\end{eqnarray}
These special analytic solutions are referred as ``fixed rays'' because the ratios of the renormalized couplings remain fixed along the flows. One immediately notices that these special set of solutions are nothing but the peculiar Ansatz found in the numerics. As explained in the introduction, if the RG potential exists, these fixed rays are the asymptotes of the ``valleys/ridges'' of the potential profile and capture the ultimate fate of RG flows completely.

After familiarizing the readers with the one-loop RG equations, we now proceed to construct the RG potential explicitly for Q1D systems. It turns out that it is convenient to represent all interacting vertices in Eqs.~(\ref{int1}) and (\ref{int2}) by Majorana fermions.

\subsection{Majorana Representation}

Now we switch to the Majorana fermion basis and construct the RG potential explicitly. Without interactions, the
band structure in low-energy limits is
described by $N_f$ flavors of Dirac fermions with {\em different} velocities in general. Each flavor of Dirac fermions can be decomposed into two Majorana fermions. Combined with spin degeneracy, the $4 N_{f}$ flavors of Majorana fermions are described by the Hamiltonian density
\begin{equation}
{\cal H}_{0} = \eta_{\mysw{R}a} (-i v_{a}\partial_x)
\eta_{\mysw{R}a} + \eta_{\mysw{L}a} (i
v_{a}\partial_x) \eta_{\mysw{L}a},
\end{equation}
where $v_a$ denotes the Fermi velocity for each flavor.

\begin{figure}
\centering
\includegraphics[width=5cm]{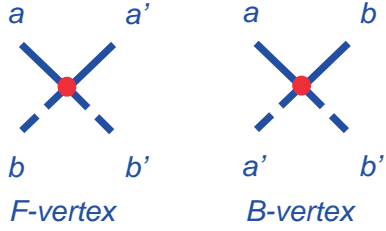}
\caption{\label{fig:Vertices} Forward and backward vertices. Bold
lines represent right-moving Majorana fermions while dashed lines
stand for the left-moving ones.}
\end{figure}

In general, a single vertex involves four different Fermi points, which generally would have four different velocities. However, while the vertices in Eq.~(\ref{int1}) and (\ref{int2}) may involve four different bands, there are at most two different velocities associated with each vertex because $v_{i}=v_{\hat{i}}$, i.e. the velocities in each vertex always appear pairwise. This seemingly useless feature turns out to be strong enough to guarantee the existence of the RG potential when the Hamitonian is re-expressed in terms of Majorana fermions. The momentum conservation in weak coupling somehow gives rise to the interesting constraint that the Fermi velocities must equal pairwise! Therefore, the interacting Hamiltonian in terms of the Majorana fermions take the form,
\begin{eqnarray}
{\cal H}_{int} &=& \tilde{F}(a,a';b,b') \eta^{}_{\mysw{R}a}
\eta^{}_{\mysw{R}a'} \eta^{}_{\mysw{L}b} \eta^{}_{\mysw{L}b'}
\nonumber\\
&+& \tilde{B}(a,b;a',b') \eta^{}_{\mysw{R}a} \eta^{}_{\mysw{R}b}
\eta^{}_{\mysw{L}a'} \eta^{}_{\mysw{L}b'},
\label{FBVertices}
\end{eqnarray}
where summations over allowed indices are implied. We emphasize that the allowed interactions might involve four different Fermi points labeled by $a,a',b,b'$, but only two different velocities $v_{a}=v_{a'}$ and $v_{b}=v_{b'}$ appear in a single vertex. By direct comparison, it is clear that the $F-$vertex include $\tilde{f}_{ij}$ and $\tilde{w}_{ij}$ and the $B-$vertex cover $\tilde{c}_{ij}$, $\tilde{s}_{ij}$ and $\tilde{u}_{ij}$.

There are also other kinds of interactions allowed by momentum
conservation,
\begin{eqnarray}
H_{c} &=&  \tilde{R}(a,a';b,b') \eta^{}_{\mysw{R}a} \eta^{}_{\mysw{R}a'}
\eta^{}_{\mysw{R}b} \eta^{}_{\mysw{R}b'} 
\nonumber\\
&+& \tilde{L}(a,a',b,b')
\eta^{}_{\mysw{L}a} \eta^{}_{\mysw{L}a'} \eta^{}_{\mysw{L}b}
\eta^{}_{\mysw{L}b'}. 
\label{ChiralVertices}
\end{eqnarray}
The scaling dimensions of these vertices are $(\Delta_{\mysw{R}},
\Delta_{\mysw{L}}) = (2,0), (0,2)$, while the vertices in
Eq.~(\ref{FBVertices}) have scaling dimensions $(1,1)$. Since the
renormalization comes from loop integrations, only vertices with
scaling dimensions differed by $(n, n)$, where $n$ is an integer,
would renormalize each other. As a result, the chiral vertices in
Eq.~(\ref{ChiralVertices}) remain marginal and only renormalize
the corresponding Fermi velocities. Since the corrections only
show up at two-loop order, we would ignore their contribution
here.  The pairwise-equal Fermi velocities in
Eq.~(\ref{FBVertices}) make the classification of all vertices
fairly simple as shown in Fig.~\ref{fig:Vertices}. The names come from the fact that the forward-type ($F$-) vertices include the usual forward scatterings while the backward-type ($B$-) vertices include the backward scatterings.

\begin{figure}
\centering
\includegraphics[width=8cm]{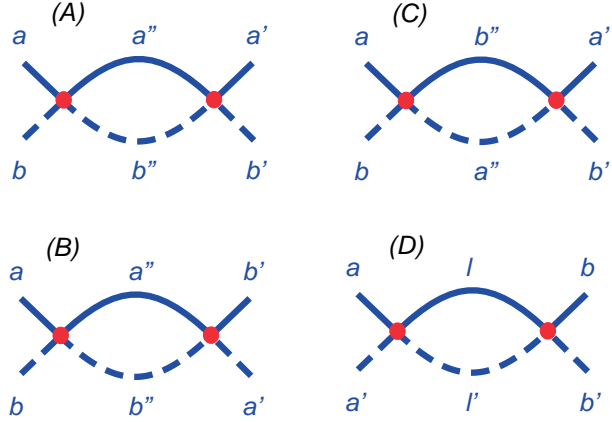}
\caption{\label{fig:Loops} Four different diagrams to the one-loop
order. (A) FF $\to$ F (B) FB $\to$ B (C)BB $\to$ F (D) BB $\to$ B.
Notice that, only in the fourth diagram, there are three
velocities involved while only two velocities are involved in all
other diagrams.}
\end{figure}

To obtain the flow equations, we need to integrate out
fluctuations at shorter length scale successively. The most
convenient approach is by the operator product expansions we introduced in previous sections. To one-loop order, the renormalization of the bare couplings come from four types of diagrams $FF \to F$, $FB \to B$, $BB \to F$ and $BB \to B$, shown in Fig.~\ref{fig:Loops}. Let us start with the first type of loop diagrams, $FF \to F$ in Fig.~\ref{fig:Loops}(a). 
The OPE of Majorana fermions can be computed
straightforwardly and the mode elimination leads to the
renormalized Hamiltonian density,
\begin{eqnarray}
\delta {\cal H}^{R} &=& \tilde{F}(a,a'';b,b'') \tilde{F}(a'',a';b'',b')
\frac{\eta^{}_{\mysw{R}a} \eta^{}_{\mysw{R}a'}
\eta^{}_{\mysw{L}b} \eta^{}_{\mysw{L}b'}}{\pi (v_a+v_b)} dl
\nonumber\\
&=&  d\tilde{F}(a,a';b,b') \eta^{}_{\mysw{R}a} \eta^{}_{\mysw{R}a'}
\eta^{}_{\mysw{L}b} \eta^{}_{\mysw{L}b'},
\end{eqnarray}
where $dl = \ln b$ is the logarithmic length scale. Here we have used the fact that $v_{a}=v_{a''}=v_{a'}$ and $v_{b}=v_{b''}=v_{b'}$. The factor
$1/\pi(v_a+v_b)$ arises from the product of propagators with opposite chiralities and different velocities. Introducing a simple rescaling of the original couplings according to their associated velocities 
\begin{equation}
F(a,a';b,b') = \frac{1}{2\pi \sqrt{v_a v_b}}
\tilde{F}(a,a';b,b'),
\end{equation}
the RG equation is
\begin{equation}
\frac{dF(a,a';b,b')}{dl} = \gamma_{ab} F(a,a'';b,b'') F(a'',a';b'',b') + ...,
\end{equation}
where $ \gamma_{ab} = 2\sqrt{v_a v_b}/(v_a +v_b)$. Repeating similar calculations, the RG equation for $F(a,a'';b,b'')$ contains a term $\gamma_{ab} F(a,a';b,a') F(a'',a';b'',b)$ and similar result for the renormalization of $F(a'',a';b'',b')$. Therefore, the flow equations can be derived from a potential,
\begin{equation}
V(F) = - \gamma_{ab} F(a,a';b,b') F(a,a'';b,b'') F(a'',a';b''b').
\end{equation}

The RG potential for the $(FBB)$ and $(BBB)$ cases shown
in Fig.~\ref{fig:Loops}(b-c) and \ref{fig:Loops}(d) can be constructed in a similar fashion. Finally, combining all contributions together, the weak-coupling RG flows for Q1D systems are described by the potential
\begin{eqnarray}
\lefteqn{V(F,B) = -B(a,b;a',b') B(b,c;b',c') B(c,a;c',a')}
\nonumber\\
&-& \gamma_{ab} B(a,b';a',b) B(a'',b';a',b'') F(a,a'';b,b'')
\nonumber\\
&-&
\gamma_{ab} F(a,a';b,b') F(a,a'';b,b'') F(a'',a';b''b'),
\label{RGPotential}
\end{eqnarray}
Summations over all allowed indices are again implied. The merits to use Majorana representation enables us to construct the RG potential explicitly. Compared with previous examples for the half-filled two-leg ladder and the doped $N$-leg system at generic filling, one can easily see that the RG potentials in Eqs.~(\ref{TwoLegRGP}) and (\ref{NLegRGP}) agree with the general form in Eq.~(\ref{RGPotential}). Note that if one starts from a `wrong' basis, it is far from trivial to realize the fact that the non-linear flows can be derived from a single potential. However, in the Majorana representation, the linear transformation to the potential basis is diagonal $L_{ij} = r_{i} \delta_{ij}$. After appropriate rescaling of couplings, the explicit form of the RG potential is derived. We have checked for the doped and half-filled Q1D systems and found all potentials agree with Eq.~(\ref{RGPotential}).

There is one loose end about the rescaling factors. For most physical systems, the rescaling factor is slightly more complicated than $(2\pi \sqrt{v_a v_b})^{-1}$. the subtlety arises from the degeneracies of the couplings imposed by physical symmetries. This is best illustrated by the following simple example. Consider the RG equations for three couplings $g_i$,
where $i=1,2,3$,
\begin{equation}
\frac{dg_{i}}{dl} = \sum_{jk} \frac{|\epsilon_{ijk}|}{2} g_{j} g_{k}.
\end{equation}
Since $|\epsilon_{ijk}|$ is totally symmetric, the corresponding
RG potential is $V(g) = g_1 g_2 g_3$. Suppose the system has some
symmetry, such as $U(1)$ symmetry for charge conservation, and the
couplings are degenerate $g_2 =g_3$. The RG equations are simplified,
\begin{equation}
\frac{dg_{1}}{dl} = g_2^2, \qquad \frac{dg_{2}}{dl} = g_1 g_2.
\end{equation}
It is straightforward to show that we need to perform a rescaling
transformation, $(h_1, h_2) = (g_1, \sqrt{2} g_2)$ to obtain the
potential $V(h) = h_1 (h_2)^2/2$. In fact, for couplings with
$n$-fold degeneracy, an additional rescaling factor $\sqrt{n}$ is
necessary to bring them into the potential basis. Therefore, the total rescaling factor is
\begin{equation}
r_{i} = \frac{1}{2\pi} \sqrt{\frac{n_i}{v_a v_b}},
\end{equation}
where $n_i$ is the degeneracy number of the coupling $g_i$, with Fermi velocities $v_a$ and $v_b$.

So far, we have shown the existence of the RG potential by explicit construction, which proves the the widely used Ansatz in Eq.~(\ref{Scaling}). In fact, the asymptotes of the RG flows are governed by the special set of fixed-ray solutions. This in turns explains the simplicity of the phase diagram, even though the RG equations are rather complicated.

The absence of exotic fates of the RG flows is also found in earlier work on 2D melting theory\cite{Kosterlitz73} or the Kondo related problems\cite{Anderson70}. However, the simplicity of RG flows in these systems is not quite the same as described here. Since the number of marginal couplings in these problems is few, analytic solution of the flows often show that it is possible to define some conserved quantity associated with each flow lines. This is where the simplicity comes from. On the other hand, we have also looked into these well-known flows to check whether they can be derived from a single potential. It is not too surprising that this is indeed the case because the requirement of potential flows loose up quite a bit when the number of couplings is small.

With the help of non-perturbative Abelian bosonization, it is not essentially important whether the RG flows can be cast into potential form beyond one-loop order. However, it remains an interesting and open question at this moment. Note that the coefficients of the one-loop RG equations are unique, protected by the leading logarithmic divergences. The next order calculations bring in lots of complications and subtleties, including the non-universal coefficients in the RG equations, velocity renormalization and so on. It is not clear at this moment whether it is even sensible to pursue the RG potential beyond one-loop order.

Another interesting issue concerns the connection between the
potential $V(h_i)$ and the Zamolodchkov's c-function $C(g_i)$ of
(1+1)-dimensional systems with Lorentz and translational
symmetries\cite{Zamolodchikov86}.  A generic Q1D system we studied here has neither Lorentz invariance (due to different Fermi
velocities) nor translation symmetry (due to Umklapp processes).
While both $V$ and $C$ are non-increasing along the RG flows, the exact
relation between them remains unclear at this point.  We emphasize
that the existence of a non-decreasing function $C$ along RG flows
only implies that $dC/dl = (\partial C/\partial g_i)\cdot
(dg_i/dl) \leq 0$ and is not strong enough to show that the flows
can be derived from a potential.  Thus, the potential flows are
closely related to the c-theorem but they are not equivalent in
general. In addition, we do not know any easy generalization of
c-theorem that does not rely on the Lorentz and translational
symmetries.  However, one can easily check that in the special
limiting case where Lorentz and translation
symmetries are restored, the $C$-function indeed coincides with the potential we
find.  This indicates that there may be a general form of
c-theorem waiting to be discovered.

In summary of this section, we have shown that the RG transformation for Q1D systems in weak coupling is described by potential flows.
Therefore, neither chaotic behaviors nor exotic limit cycles could
occur. The different Fermi velocities and the degeneracies
imposed by physical symmetries give rise to non-trivial rescaling
factors, which hinder this beautiful structure behind the RG
transformation. The explicit form of the potential is obtained
after appropriate rescaling of the couplings in Majorana basis.

\section{Anomalous Scaling}

Now that we have established the existence of the RG potential, we would like to discuss its physical consequences. Since the notion of RG potential is still novel to the community of low-dimensional correlated physics, we concentrate on the simplest example, the half-filled two-leg ladder, in this section. However, one should keep in mind that most of the results discussed here can be applied to more general ladder systems.

For the two-leg ladder at half filling, the number of possible 
interactions is greatly reduced to nine in weak coupling. Within the 
one-loop RG calculations, these couplings $g_{i}$ are described 
by a set of coupled first-order differential equations as in Eq.~(\ref{OneLoop}). For the two-leg ladder at half filling, $M^{i}_{jk}$ are $9 \times 9$ constant matrices, as implicitly contained in the RG equations in Appendix B. For a generic interacting Hamiltonian, the bare 
values for these nine couplings can be straightforwardly determined. 
However, it is generally very difficult to obtain the solution for 
these coupled flow equations in analytical form.

Simple analytical solutions emerge if the interactions are chosen in a 
specific way. These special solutions are later referred as 
``symmetric rays''. Suppose the bare couplings of the specific 
interacting Hamiltonian are $ g_{i}(0) = G_{i} g(0)$, where $g(0) = 
(U/t) \ll 1$ is small while $G_{i}$ are order one constants which 
satisfy the algebraic constraint,
\begin{equation}
G_{i} = M^{i}_{jk} G_{j} G_{k}.	
\end{equation}
It is straightforward to show that the ratios between couplings 
remain the same and the nine complicated equations reduce to single 
one, $\dot{g} = g^{2}$ ! The solution is $g(l)=1/(l_{d}-l)$, where 
the divergent length scale $l_{d} = (t/U)$. Of course, one should 
keep in mind that the solution $g(l)$ is only valid when it does not 
flow out of the weak coupling regime. These special Hamiltonians, 
whose couplings are described by these symmetric rays, turn out to be 
SO(8) symmetric. For the two-leg ladder, four different phases\cite{Lin98,Fjaerestad02}, named as D-Mott, S-Mott, CDW and SF, are of the central concerns.

It was shown previously that the two-leg ladder in weak coupling {\it 
always} scales into one of the four different symmetric phases 
\cite{Lin98}. However, this numerical approach was criticized that the 
SO(8) symmetric phases might have instabilities which happen not to be 
tackled by the limited types of interactions considered in the 
numerical study. To make up the fissure, a complete stability check 
near the SO(8) symmetric rays is desirable.

\subsection{Stability Analysis}

To describe the RG flows in the vicinity of the SO(8) symmetric rays, 
it is sufficient to consider the linearized version of Eq.~(\ref{OneLoop}). 
For a generic interaction, the couplings are separated into symmetric 
and asymmetric parts, $g_{i}(l) = G_{i} g(l) + \Delta g_{i}(l)$. In 
the vicinity of the symmetric rays, the deviations are small, $\Delta 
g_{i}(l) \ll g(l)$. Keeping the leading order term, the linearized RG 
equations are
\begin{equation}
\frac{d (\Delta g_{i})}{dl} = \frac{B_{ij}}{(l_{d}-l)} \Delta g_{i},
\end{equation}
where $B_{ij} = 2 A^{i}_{jk} r_{k}$. The matrix $B_{ij}$ can be 
brought into diagonal form by a linear transformation. As a 
consequence, the RG equations decouple into nine independent ones,
\begin{equation}
\frac{d (\delta g_{i})}{dl} = \frac{\lambda_{i}}{(l_{d}-l)} \delta g_{i},
\label{LinearRG}
\end{equation}
where $\delta g_{i}$ are couplings after the linear transformation 
and $\lambda_{i}$ are the eigenvalues of the matrix $B_{ij}$. Although 
the matrix $B_{ij}$ are different for each SO(8) symmetric rays, the 
eigenvalues are identically the same
\begin{equation}
\lambda_{i} = 2, \frac23, \frac23, \frac23, 
-\frac13, -\frac13, -\frac13, -\frac13, -\frac13.
\end{equation}
This coincidence implies that the results of the stability check only 
rely on the symmetry group but not on the details of the phases 
\cite{Konik02}.

So far, we have a single equation, $\dot{g}=g^{2}$, describing the
renormalization along the symmetric rays and nine for deviations from
the rays as in Eq.~(\ref{LinearRG}).  Apparently, there must be one
redundant equation among them because the original number of equations
is only nine.  It doesn't take long to find out that it corresponds to
the flow equation with largest eigenvalue $\lambda=2$. This 
corresponds to a trivial case that all couplings are shifted along the 
symmetric ray, i.e. $\delta g_{i} = G_{i} \delta g$. Since the flow 
along the ray is described by $\dot{g}=g^{2}$, linearization leads 
to $\delta \dot{g}= 2g(l) \delta g$. Since we only need eight 
equations to describe the deviations from the symmetric rays, the 
$\lambda=2$ equation is only an artifact and should be ignored.

\begin{figure}
\centering
\includegraphics[height=3cm]{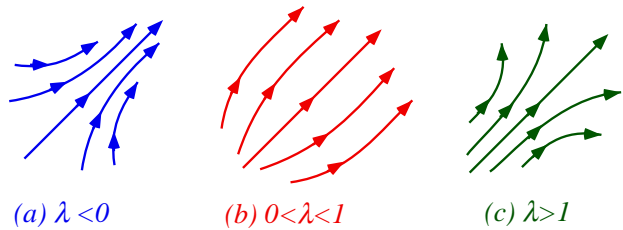}
\caption{\label{Relevance}The topology of RG flows near the symmetric ray with (a) $\lambda<0$, (b) $0<\lambda<1$ and (c) $\lambda>1$. It is clear that the coupling is irrelevant for $\lambda<0$ and relevant for 
$\lambda>1$. The RG flow is more subtle for $0<\lambda<1$. In this 
case, although the deviation from the symmetric ray is growing, the 
slope remains the same as the ray.}
\end{figure}

The other eight eigenvalues describe how the flow goes once the bare 
couplings are off the ray. Starting from the bare values $\delta 
g_{i}(0) = \delta G_{i} g(0)$, where $\delta G_{i} \ll G_{i}$. The 
solutions of Eq.~(\ref{LinearRG}) are
\begin{equation}
\delta g_{i}(l) = \frac{\delta G_{i}}{(l_{d}-l)^{\lambda_{i}}} 
\left(\frac{U}{t} \right)^{1-\lambda_{i}}.	
\label{Deviations}
\end{equation}
According to conventional classification \cite{Goldenfeld92}, for 
$\lambda_{i}<0$ the deviations diminish under RG transformations as 
shown in Fig.~\ref{Relevance}(a) and thus are classified as irrelevant couplings. For $\lambda_{i}>0$, any small deviations get enhanced and the flow is pushed away from the symmetric rays as in Fig.~\ref{Relevance}(b) and \ref{Relevance}(c). 
These are classified as relevant couplings. The conventional wisdom tells us 
that there are three relevant couplings ($\lambda_{i} = 2/3$) and five 
irrelevant ones ($\lambda_{i}=-1/3$). Since a generic interaction in 
principle could generate asymmetric deviations in all couplings, one 
might rush to the incorrect conclusion that the SO(8) symmetry is not 
stable. However, the first-glance guess is wrong because the 
conventional classification is based on the perturbative analysis 
near a {\em fixed point} while we are dealing with running symmetric rays. 
A new set of rules to identify relevant perturbations is in order.

The crucial criterion is whether the deviations $\delta g_{i}(l)$ grow
larger than the symmetric coupling $g(l)=(l_{d}-l)^{-1}$.  Just
growing larger than the bare values under RG transformations is not
qualified as a relevant perturbation. This subtle but important 
difference is best illustrated by calculating gap functions using 
scaling arguments. If the degeneracy of excitation gaps is 
maintained, the SO(8) symmetry is robust and vice versa.

\subsection{Scaling Argument and Anomalous Exponents}

Now we elaborate on the anomalous scaling arisen from the peculiar RG flows near the symmetric rays. Let us remind you that there are several useful properties of the RG flows in Eq.~(\ref{OneLoop}). Suppose all bare couplings are rescaled by a factor of $U$, $\tilde{g}_{i}(0) = U g_{i}(0)$, it is easy to check that the RG flows are recovered by a rescaled (logarithmic) length scale $\tilde{l} = l/U$. That is to say,
\begin{equation}
\tilde{g}_{i}(l) = U g_{i}(Ul).
\end{equation}
It means that, as long as the relative ratios of the bare couplings are fixed, the RG flows for different interaction strength are related in a simple way.

To gain some physical intuition of the scaling argument, it is illuminating to study the simplest one-loop equation
\begin{equation}
\frac{dg}{dl} = -g^2,
\end{equation}
with solution $g(l) = 1/(l-l_d)$, where the divergent length scale $l_d = -1/g(0)$. The relevance of the marginal coupling $g$ depends on the sign of the bare coupling value. For $g(0)>0$, $l_d$ is negative so that the renormalized coupling $g(l)$ scale toward zero as power law. For $g(0)<0$, $g(l)$ grows under RG transformation and eventually becomes divergent at $l = l_d$. This simple RG equation actually describes the well-known BCS instability. For repulsive interaction, corresponding to $g(0)>0$, the pairing interaction is marginally irrelevant and the Fermi liquid is stable with inclusion of electronic correlations. On the other hand, for attractive interaction where $g(0)<0$, the pairing instability is enhanced and leads to superconducting ground state.

The analytical solution in the previous example also help us to understand the peculiar scaling of the gap function generated by marginally relevant couplings. We must emphasize that the appearance of a relevant coupling does not necessarily mean the formation of some excitation gap in the energy spectrum. However, let us  concentrate on the case where the relevant coupling does drive the formation of some gap. From conventional scaling argument, $\Delta[g(l)] = e^l \Delta[g(0)]$, we cut off the flow when the relevant coupling reaches one $g(l_c)=1$ because the one-loop RG equation is no longer reliable. It is easy to solve for the cutoff length scale,
$l_c = l_d-1 \approx 1/|g(0)|$. At the cutoff length scale, since the coupling is order of unity, the gap function is also order one. Thus, the scaling argument tells us that
\begin{equation}
\Delta[g(0)] = e^{-l_c} \Delta[g(l_c)] \approx \Delta_{0} \times e^{-1/|g(0)|},
\end{equation}
where $\Delta_{0}$ is some is some energy scale of the system (like the band width $t$). Although the non-analytical gap function is derived in this extremely simple case, it is expected that the gap function would behave in a similar way for more general quasi-1D systems.

Since the RG equations are only valid in weak coupling, we cut off 
the RG procedure when the symmetric coupling $g(l_{c})=1$. At this 
cutoff length scale $l_{c}$, the deviations in Eq.~(\ref{Deviations}) 
are
\begin{equation}
\delta g_{i} (l_{c}) \sim \left( \frac{U}{t}\right)^{1-\lambda_{i}}.
\end{equation}
For $\lambda_{i}>1$, the deviations are larger than the symmetric 
coupling and should be classified as ``relevant''. As long as 
$\lambda_{i}<1$, the deviations at the cutoff length scale are still 
vanishingly small and should be viewed as ``irrelevant''. This new 
classification is different from the conventional one for 
$0<\lambda_{i}<1$. We would see clearly soon that the new 
classification is appropriate for stability check near a running 
symmetric ray. The grey regime $0<\lambda_{i}<1$ between the new and 
conventional rules gives rise to anomalous scaling exponents.

Under RG transformations, the gap functions scale like,
\begin{equation}
\Delta_{i} \left[ g(0), \delta g_{i}(0) \right] =
e^{-l_{c}} \Delta_{i} \left[ 1, \delta g_{i}(l_{c}) \right],
\end{equation}
where $\delta g_{i}(l_{c})$ are at most of order $(U/t)^{1/3}$. The 
key point is that, although some perturbations are enhanced to 
order $(U/t)^{1/3}$, which is larger than the bare order $U/t$ 
values, they are still small at the cutoff length scale. The 
effective Hamiltonian can be separated into two parts $H = H_{0} + 
\delta H$ -- the SO(8) symmetric and asymmetric parts. Since the 
asymmetric part is of order $(U/t)^{1/3}$, the changes of the gaps 
would be of the same order by standard perturbation theory. Without 
the deviations, the SO(8) symmetry guarantees the exact degeneracy of 
all gaps, i.e. $\Delta[1,0] = \Delta$. The presence of perturbations 
modifies the gap functions,
\begin{equation}
\Delta_{i}\left[1,\delta g_{i}(l_{c}) \right] = \Delta \left[
1+c_{i} \left(\frac{U}{t}\right)^{\frac13} + \ldots \right].
\end{equation}
It is clear that, in the weak coupling limit $U/t \to 0$, the 
degeneracy of all gaps is recovered. It implies that the SO(8) 
symmetry is indeed robust under generic perturbations.

\begin{figure}[htb]
\centering
\includegraphics[height=4cm]{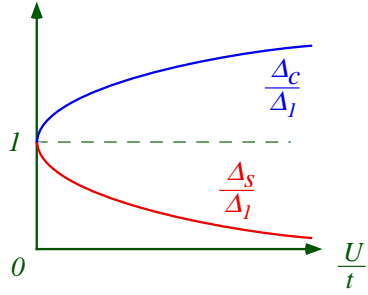}
\caption{\label{GapRatios} Ratios of charge, spin and single particle gaps plotted versus the interaction strength $U/t$. The ratios approach unity in 
the asymptotic $ U/t \to 0$ limit with anomalous corrections of 
order $(U/t)^{1/3}$.}
\end{figure}

The anomalous scaling exponent $1/3$ is clearly seen in the ratios 
between charge, spin and single particle gaps (see Fig.~\ref{GapRatios}),
\begin{equation}
\frac{\Delta_{i}}{\Delta_{j}} \approx 1 + c_{ij} \left( \frac{U}{t} 
\right)^{\frac13}.
\end{equation}
This simple exponent is rather fascinating because the RG equations we 
started from are very messy. But, despite of which SO(8) symmetric 
phases the system flows into, the exponent of the gap function 
corrections is universally equal to $1-\lambda_{i} = 1/3$!

It is quite exciting to learn that the existence of RG potential, not only explain the simplicity of the phase diagrams from the complicated RG equations, it also predicts the emergence of the enhanced symmetry in Q1D ladders. Furthermore, the instability check in the vicinity of these symmetric rays gives anomalous scaling with ratios numbers associated with the emergent symmetry. While we only worked out the details for the half-filled two-leg ladder in this section, it is expected that the phenomena should be rather general. There is still plenty of room to investigate the fun physics associated with the RG potential in Q1D ladders.

\section{Discussions and Conclusions}

In this article, we reviewed how to write down the effective action for Q1D systems in weak coupling. In contrast to conventional $g$-ology, we classify and express all possible four-fermion interactions by SU(2) bilinear currents. With the help of OPE's, the derivation of RG equations becomes equivalent to obtain the coefficients of algebra among the currents. To our best knowledge, this is the first time that the whole set of RG equations for generic Q1D systems is written down explicitly.

While numerical elaboration of the RG equations is important to reveal physical properties in the ladder materials, we focus on the hidden structure behind the scene and hope to answer two fundamental questions: why the phase diagram is so simple despite of the complexity of the non-linear RG equations and why the enhanced symmetry emerges after integrating out the short-range fluctuations. Both puzzles can be answered by the non-trivial existence of the so-called RG potential which allows a simple geometric interpretation of the flows. The ultimate destinies, after successively integrating out fluctuations at shorter length scale, are nothing but the well-isolated asymptotic ``valleys/ridges'' with specific ratios between the renormalized couplings. The isolation of the asymptotes explains the simple and robust structure of the ground-state phase diagram and the specific ratios give rise to the enhanced symmetry emerging in low-energy limit. We first demonstrated how the RG potential can be constructed explicitly in the half-filled two-leg ladder and also the doped $N$-leg ladder. Then, we provide a general proof for its existence for generic Q1D systems and also pin down the functional form of the potential.

There are many open questions left behind the work here. For instance, it is interesting to filling in the numerical details for specific ladder materials. But, maybe one of the most interesting questions is its applicability to strongly correlated systems in two dimensions. First of all, one may naively expect Q1D systems with large $N$ should share a similar ground state for a truly two-dimensional system. Unfortunately, if the Fermi surface is not nested, the instability in the RG equations fade away and the ground state is the normal Fermi liquid. The failure to deliver an exotic ground state in the presence of featureless Fermi surface may indicate two possibile scenarios. 

The first scenario is that all interesting ground states in two dimensions only occur in strong coupling and the RG analysis in weak coupling provides null information about them. If so, one should abandon the weak-coupling approach completely and search for alternative routes. On the other hand, the second scenario to explain the failure is tied up with the shape of Fermi surface. It is possible that the exotic phases in two dimensions are driven by the enhanced couplings due to nested Fermi surface. Therefore, by cutting up the Fermi surface appropriately, the RG technique developed here can be applied to determine the ground states in two dimensions. Even if the instabilities towards some exotic phases can be spotted in weak-coupling RG, it doesn't mean that we understand or know how to appropriately describe the physical properties which are essentially described by the fixed points in strong coupling. However, the powerful technique of bosonization may help us out here to sketch the ground state qualitatively. It is not clear at this stage which scenario is correct and more numerical work needs to be done to clear up the relevance between the Q1D and the true 2D physics. If the truth favors the second scenario, the RG potential would give us more confidence for what is found in weak-coupling analysis.

\section*{Acknowledgements}
We thank Leon Balents, Darwin Chang, Matthew Fisher, Y.-C. Kao, Andreas Ludwig, Chung-Yu Mou and S.-K. Yip for fruitful discussions. In particular, HHL is thankful for illuminating counter example in dynamical systems provided by Leon Balents. HHL appreciates financial supports from National Science Council in Taiwan through Ta-You Wu Fellow and grants NSC-93-2120-M-007-005 and NSC-94-2112-M-007-031. The hospitality of KITP, where part of the work was carried out, and supports from NSF PHY99-07949 are great appreciated.

\appendix
\section{Operator Product Expansions}

Following similar calculations presented in the text, one can obtain the full set of the OPE's. For interested readers who would like to derive the identities on their own, you will find the following identities useful,
\begin{eqnarray}
\epsilon_{\alpha \beta} \sigma _{\beta \gamma }^{a}
\epsilon_{\gamma \delta} = \sigma _{\delta \alpha }^{a},
\qquad
\epsilon_{\alpha \beta } \sigma _{\beta \gamma}^{a}
\sigma _{\alpha \varepsilon}^{b} = \epsilon_{\gamma \eta
}(\delta _{\eta \varepsilon}+i\varepsilon^{abc}\sigma _{\eta \varepsilon
}^{c}).
\end{eqnarray}
In below, we list the complete rules for the OPE of right-moving currents,
\begin{eqnarray}
J_{ij}(x,\tau )J_{kl}(0,0) &\sim&
\frac{1}{2 z_{i}z_{j}}\delta _{il}\delta_{jk}
+ \bigg\{\frac{\delta _{jk}}{2z_{j}}J_{il}
-\frac{\delta _{il}}{2z_{i}}J_{kj}\bigg\},
\\
J_{ij}^{a}(x,\tau )J_{kl}(0,0)&\sim&
\frac{\delta _{jk}}{2 z_{j}}J_{il}^{a}
-\frac{\delta _{il}}{2 z_{i}}J_{kj}^{a},
\\
J_{ij}^{a}(x,\tau )J_{kl}^{b}(0,0) &\sim&
\frac{\delta ^{ab}}{2z_{i}z_{j}}\delta _{il}\delta _{jk}
+\delta ^{ab}\bigg\{
\frac{\delta _{jk}}{2z_{j}}J_{il}
-\frac{\delta _{il}}{2z_{i}}J_{kj}\bigg\}
\nonumber\\
&& + i\epsilon ^{abc}\bigg\{
\frac{\delta _{jk}}{2 z_{j}}J_{il}^{c}
+\frac{\delta _{il}}{2 z_{i}}J_{kj}^{c}\bigg\},
\end{eqnarray}
\begin{eqnarray}
I_{ij}^{\dag }(x,\tau )I_{kl}^{\vphantom\dag }(0,0)\!&\sim&\! 
\frac{1}{2 z_{i}z_{j}}(\delta _{ik}\delta _{jl}\!+\!\delta _{il}\delta _{jk})
\! + \!\! \bigg\{
\frac{\delta _{ik}}{2 z_{i}}J_{jl}+\frac{\delta _{il}}{2 z_{i}}J_{jk}
+\frac{\delta _{jk}}{2 z_{j}}J_{il}+\frac{\delta _{jl}}{2 z_{j}}J_{ik}\bigg\},
\\
I_{ij}^{a\dag }(x,\tau )I_{kl}^{\vphantom\dag }(0,0) &\sim& 
\frac{\delta _{ik}}{2 z_{i}}J_{jl}^{a}+\frac{\delta _{il}}{2 z_{i}}J_{jk}^{a}
-\frac{\delta _{jk}}{2 z_{j}}J_{il}^{a}-\frac{\delta _{jl}}{2 z_{j}}J_{ik}^{a},
\\
I_{ij}^{a\dag }(x,\tau )I_{kl}^{b}(0,0) &\sim& 
\frac{\delta ^{ab}}{2z_{i}z_{j}}(\delta _{ik}\delta _{jl}-\delta _{il}\delta _{jk})
+\delta ^{ab} \bigg\{\frac{\delta _{ik}}{2 z_{i}}J_{jl}
-\frac{\delta _{il}}{2 z_{i}}J_{jk}-\frac{\delta _{jk}}{2 z_{j}}J_{il}
+\frac{\delta_{jl}}{2 z_{j}}J_{ik}\bigg\}
\nonumber\\
&& +i\epsilon ^{abc} \bigg\{
\frac{\delta _{ik}}{2 z_{i}}J_{jl}^{c}
-\frac{\delta _{il}}{2 z_{i}}J_{jk}^{c}
-\frac{\delta _{jk}}{2 z_{j}}J_{il}^{c}
+\frac{\delta _{jl}}{2 z_{j}}J_{ik}^{c}\bigg\},
\end{eqnarray}
\begin{eqnarray}
I_{ij}(x,\tau )J_{kl}(0,0) &\sim &
\frac{\delta _{ik}}{2 z_{i}}I_{jl}
+\frac{\delta _{jk}}{2 z_{j}}I_{il},
\label{Ir_Jr} \\
I_{ij}(x,\tau )J_{kl}^{a}(0,0) &\sim &
\frac{\delta _{ik}}{2 z_{i}}I_{jl}^{a}
+\frac{\delta _{jk}}{2 z_{j}}I_{il}^{a},
\\
I_{ij}^{a}(x,\tau )J_{kl}(0,0) &\sim &
-\frac{\delta _{ik}}{2 z_{i}}I_{jl}^{a}
+\frac{\delta _{jk}}{2 z_{j}}I_{il}^{a},
\\
I_{ij}^{a}(x,\tau )J_{kl}^{b}(0,0) &\sim &
\delta ^{ab}\bigg\{\!\!-\frac{\delta _{ik}}{2 z_{i}}I_{jl}
+\frac{\delta _{jk}}{2 z_{j}}I_{il}\bigg\}
+i\epsilon ^{abc}\bigg\{\!\!
-\frac{\delta _{ik}}{2 z_{i}}I_{jl}^{c}
+\frac{\delta _{jk}}{2 z_{j}}I_{il}^{c}\bigg\}.
\end{eqnarray}
There are other OPE's, for example, $I_{ij}^{\dag }J_{kl}$, can be obtained
by taking hermitian conjugate in Eq.~(\ref{Ir_Jr}) on both sides and sending $z_{i}\to -z_{i}$ at the same time. Replacing $z_{i}\to z_{i}^{*}$ in the
above equations gives the OPE's for left moving currents.

\section{RG Equations}

Since the four-fermion interactions in Q1D system are all marginal, the renormalized couplings are described by coupled non-linear equations. In addition to the usual forward and Cooper scatterings, denoted as $f^{\rho,\sigma}_{ij}$ and $c^{\rho,\sigma}_{ij}$, the nested Fermi surface also host the extra momentum-conserving vertices $s^{\rho,\sigma}_{ij}$ and the umklapp interactions $u^{\rho,\sigma}_{ij}$, $w^{\rho,\sigma}_{ij}$. After integrating out fluctuations at short-length scale, the complete RG equations for the Q1D system at half filling are shown in the followings:

\begin{eqnarray}
{\dot{c}}_{ij}^{\rho } &=&-\sum_{k}\frac{\alpha _{ij,k}}{4}
\bigg\{
c_{ik}^{\rho }c_{kj}^{\rho }+3c_{ik}^{\sigma }c_{kj}^{\sigma }
\bigg\}
+\frac{1}{4}(c_{ij}^{\rho }h_{ij}^{\rho }+3c_{ij}^{\sigma }h_{ij}^{\sigma })
\nonumber \\
&+&\sum_{k}\delta _{i{\hat{j}}}\frac{\alpha _{ii,k}}{4}
\bigg\{
s_{ik}^{\rho }s_{k{\hat{i}}}^{\rho}
+3s_{ik}^{\sigma }s_{k{\hat{i}}}^{\sigma}
+ u_{ik}^{\rho }u_{k{\hat{i}}}^{\rho}
+ 3u_{ik}^{\sigma }u_{k{\hat{i}}}^{\sigma }
\bigg\}  
\nonumber \\
&+&\frac{\delta _{ij}}{4}
\bigg\{ 
(u_{i{\hat{i}}}^{\rho })^{2}+3(u_{i{\hat{i}}}^{\sigma })^{2}
\bigg\}
+\frac{\delta _{i{\hat{j}}}}{2}(u_{ii}^{\rho }u_{i{\hat{i}}}^{\rho}) +(u_{i{\hat{j}}}^{\rho}w_{ij}^{\rho }
-3u_{i{\hat{j}}}^{\sigma }w_{ij}^{\sigma }),
\end{eqnarray}

\begin{eqnarray}
{\dot{c}}_{ij}^{\sigma} &=&-\sum_{k}\frac{\alpha _{ij,k}}{4}
\bigg\{
c_{ik}^{\rho }c_{kj}^{\sigma}
+ c_{ik}^{\sigma}c_{kj}^{\rho}
+ 2c_{ik}^{\sigma }c_{kj}^{\sigma}
\bigg\}
+ \frac{1}{4}(c_{ij}^{\rho}h_{ij}^{\sigma}
+ c_{ij}^{\sigma}h_{ij}^{\rho }
- 2c_{ij}^{\sigma}h_{ij}^{\sigma})
\nonumber\\
&& +\sum_{k}\delta _{i{\hat{j}}}\frac{\alpha _{ii,k}}{4}
\bigg\{
(s_{ik}^{\rho}s_{k{\hat{i}}}^{\sigma}
+ s_{ik}^{\sigma }s_{k{\hat{i}}}^{\rho }
- 2s_{ik}^{\sigma }s_{k{\hat{i}}}^{\sigma })
+ (u_{ik}^{\rho}u_{k{\hat{i}}}^{\sigma}
+ u_{ik}^{\sigma}u_{k{\hat{i}}}^{\rho}
- 2u_{ik}^{\sigma}u_{k{\hat{i}}}^{\sigma})
\bigg\}
\nonumber\\
&&+\frac{\delta_{i{\hat{j}}}}{2}u_{ii}^{\rho} u_{i{\hat{i}}}^{\sigma}
- \frac{\delta_{ij}}{2}
\bigg\{
u_{i{\hat{i}}}^{\rho}u_{i{\hat{i}}}^{\sigma}
+ (u_{i{\hat{i}}}^{\sigma})^{2}
\bigg\}
+(u_{i{\hat{j}}}^{\rho}w_{ij}^{\sigma}
-u_{i{\hat{j}}}^{\sigma}w_{ij}^{\rho}
+ 2u_{i{\hat{j}}}^{\sigma}w_{ij}^{\sigma}),
\end{eqnarray}

\begin{eqnarray}
{\dot{f}}_{ij}^{\rho } &=&
\frac{1}{4}\left[ (c_{ij}^{\rho})^{2}+3(c_{ij}^{\sigma})^{2}\right]  
\nonumber\\
&+& \sum_{k}\delta _{i{\hat{j}}}\frac{\alpha _{ii,k}}{4}
\bigg\{
(s_{ik}^{\rho})^{2}+3(s_{ik}^{\sigma})^{2}
+ (u_{ik}^{\rho})^{2}+3(u_{ik}^{\sigma})^{2}
\bigg\}  
\nonumber\\
&+&\frac{\delta _{i{\hat{j}}}}{2}(u_{ii}^{\rho})^{2}
-\frac{1}{4}[(s_{i{\hat{j}}}^{\rho})^{2}
+ 3s_{i{\hat{j}}}^{\sigma})^{2}]
+ \frac{1}{4}[(u_{i{\hat{j}}}^{\rho})^{2}
+ 3(u_{i{\hat{j}}}^{\sigma})^{2}]
+[(w_{ij}^{\rho})^{2}+3(w_{ij}^{\sigma})^{2}],
\nonumber\\
&&
\end{eqnarray}

\begin{eqnarray}
{\dot{f}}_{ij}^{\sigma } &=&
-(f_{ij}^{\sigma})^{2} 
+ \frac{1}{2}c_{ij}^{\rho}c_{ij}^{\sigma}
- \frac{1}{2}(c_{ij}^{\sigma})^{2}  
\nonumber\\
&+&\sum_{k}\delta_{i{\hat{j}}}\frac{\alpha_{ii,k}}{2}
\bigg\{
s_{ik}^{\rho}s_{ik}^{\sigma}
- (s_{ik}^{\sigma})^{2}
+ u_{ik}^{\rho}u_{ik}^{\sigma}
- (u_{ik}^{\sigma })^{2}]
\bigg\}  
\nonumber\\
&+&2 [w_{ij}^{\rho}w_{ij}^{\sigma}
- (w_{ij}^{\sigma })^{2}]
- \frac{1}{2}[u_{i{\hat{j}}}^{\rho}u_{i{\hat{j}}}^{\sigma}
+ (u_{i{\hat{j}}}^{\sigma})^{2}
+ s_{i{\hat{j}}}^{\rho}s_{i{\hat{j}}}^{\sigma}
+ (s_{i{\hat{j}}}^{\sigma})^{2}]
\end{eqnarray}

\begin{eqnarray}
{\dot{s}}_{ij}^{\rho } &=&\sum_{k}\frac{\alpha_{ij,k}}{4}
\bigg\{
s_{ik}^{\rho}s_{kj}^{\rho}
+ 3s_{ik}^{\sigma}s_{kj}^{\sigma}
+ u_{ik}^{\rho}u_{kj}^{\rho}
+ 3u_{ik}^{\sigma}u_{kj}^{\sigma}
\bigg\}  
\nonumber \\
&+&\frac{1}{4}(u_{ij}^{\rho}u_{ij}^{\rho 0}
+ s_{ij}^{\rho}f_{ij}^{\rho-}
+ 3s_{ij}^{\sigma}f_{ij}^{\sigma-}
+ s_{i{\hat{j}}}^{\rho}c_{ij}^{\rho 0}
+ 3 s_{i{\hat{j}}}^{\sigma}c_{ij}^{\sigma 0}),
\end{eqnarray}

\begin{eqnarray}
{\dot{s}}_{ij}^{\sigma} &=&\sum_{k}\frac{\alpha_{ij,k}}{4}
\bigg\{
s_{ik}^{\rho}s_{kj}^{\sigma}
+ s_{ik}^{\sigma}s_{kj}^{\rho}
- 2s_{ik}^{\sigma}s_{kj}^{\sigma}
+ u_{ik}^{\rho}u_{kj}^{\sigma}
+ u_{ik}^{\sigma}u_{kj}^{\rho}
- 2u_{ik}^{\sigma}u_{kj}^{\sigma}
\bigg\}
\nonumber \\
&+&\frac{1}{4}(s_{i{\hat{j}}}^{\rho}c_{ij}^{\sigma 0}
+ s_{i{\hat{j}}}^{\sigma} c_{ij}^{\rho 0}
- 2 s_{i{\hat{j}}}^{\sigma}c_{ij}^{\sigma 0}
+ u_{ij}^{\sigma}u_{ij}^{\rho 0})
+ \frac{1}{4}(s_{ij}^{\rho}f_{ij}^{\sigma -}
+ s_{ij}^{\sigma}f_{ij}^{\rho -}
- 2 s_{ij}^{\sigma}f_{ij}^{\sigma +}),
\nonumber \\
&&
\end{eqnarray}

\begin{eqnarray}
{\dot{u}}_{ij}^{\rho} &=&\sum_{k}\frac{\alpha _{ij,k}}{4}
\bigg\{
(u_{ik}^{\rho}s_{kj}^{\rho}
+ s_{ik}^{\rho}u_{kj}^{\rho})
+ 3(u_{ik}^{\sigma}s_{kj}^{\sigma}
+ s_{ik}^{\sigma}u_{kj}^{\sigma})
\bigg\}  
\nonumber \\
&&+\frac{\delta_{i{\hat{j}}}}{2}
\bigg\{
u_{ii}^{\rho}c_{i{\hat{i}}}^{\rho}
+ u_{i{\hat{i}}}^{\rho}c_{ii}^{\rho}
- \delta _{ij}u_{ii}^{\rho}c_{ii}^{\rho}
- 3u_{i{\hat{i}}}^{\sigma}c_{ii}^{\sigma}
\bigg\}
\nonumber\\
&&+\frac{1}{4}(u_{i{\hat{j}}}^{\rho}c_{ij}^{\rho 0}
+3 u_{i{\hat{j}}}^{\sigma} c_{ij}^{\sigma 0}
+ u_{ij}^{\rho } f_{ij}^{\rho +}
+ 3 u_{ij}^{\sigma} f_{ij}^{\sigma -}
+ u_{ij}^{\rho 0} s_{ij}^{\rho})
\nonumber\\
&&+ (w_{ij}^{\rho}c_{i{\hat{j}}}^{\rho}
- 3 w_{ij}^{\sigma}c_{i{\hat{j}}}^{\sigma}),
\end{eqnarray}

\begin{eqnarray}
{\dot{u}}_{ij}^{\sigma} &=&\sum_{k}\frac{\alpha _{ij,k}}{4}
\bigg\{
(u_{ik}^{\rho}s_{kj}^{\sigma}
+ s_{ik}^{\sigma}u_{kj}^{\rho}
+ u_{ik}^{\sigma}s_{kj}^{\rho}
+ s_{ik}^{\rho}u_{kj}^{\sigma}
- 2 s_{ik}^{\sigma}u_{kj}^{\sigma}
- 2 u_{ik}^{\sigma}s_{kj}^{\sigma})
\bigg\}
\nonumber \\
&&+\frac{\delta_{i{\hat{j}}}}{2}(u_{ii}^{\rho}c_{i{\hat{i}}}^{\sigma}
- u_{i{\hat{i}}}^{\rho}c_{ii}^{\sigma}
+ u_{i{\hat{i}}}^{\sigma} c_{ii}^{\rho}
- 2 u_{i{\hat{i}}}^{\sigma}c_{ii}^{\sigma})
+ \frac{1}{4}(u_{i{\hat{j}}}^{\rho}c_{ij}^{\sigma 0}
+ u_{i{\hat{j}}}^{\sigma}c_{ij}^{\rho 0}
- 2 u_{i{\hat{j}}}^{\sigma}c_{ij}^{\sigma 0})
\nonumber \\
&&+\frac{1}{4}(u_{ij}^{\rho}f_{ij}^{\sigma -}
+ u_{ij}^{\sigma} f_{ij}^{\rho +}
- 2 u_{ij}^{\sigma} f_{ij}^{\sigma +}
+ u_{ij}^{\rho 0} s_{ij}^{\sigma})
+ (c_{i{\hat{j}}}^{\rho}w_{ij}^{\sigma}
- c_{i{\hat{j}}}^{\sigma}w_{ij}^{\rho}
- 2 c_{i{\hat{j}}}^{\sigma}w_{ij}^{\sigma}),
\nonumber\\
&&
\end{eqnarray}

\begin{eqnarray}
{\dot{w}}_{ij}^{\rho} &=&\frac{1}{2}(u_{i{\hat{j}}}^{\rho}c_{ij}^{\rho}
- 3u_{i{\hat{j}}}^{\sigma}c_{ij}^{\sigma})
\nonumber \\
&+&\frac{1}{2}\bigg\{ 
w_{ij}^{\rho}(f_{ij}^{\rho}+f_{i{\hat{j}}}^{\rho})
+ 3w_{ij}^{\sigma}(f_{ij}^{\sigma}-f_{i{\hat{j}}}^{\sigma})
\bigg\},
\end{eqnarray}

\begin{eqnarray}
{\dot{w}}_{ij}^{\sigma} &=& \frac{1}{2}(u_{i{\hat{j}}}^{\rho}c_{ij}^{\sigma}
- u_{i{\hat{j}}}^{\sigma}c_{ij}^{\rho}
+ 2u_{i{\hat{j}}}^{\sigma}c_{ij}^{\sigma})
\nonumber \\
&+&\frac{1}{2}\bigg\{ 
w_{ij}^{\rho}(f_{ij}^{\sigma}
- f_{i{\hat{j}}}^{\sigma})
+ w_{ij}^{\sigma}(f_{ij}^{\rho}+f_{i{\hat{j}}}^{\rho})
- 2w_{ij}^{\sigma}(f_{ij}^{\sigma}+f_{i{\hat{j}}}^{\sigma})
\bigg\}.
\end{eqnarray}
Notice all couplings in the RG equations are rescaled with a velocity factor to simplify the coefficients, $g_{ij}\equiv \tilde{g}_{ij}/\pi (v_{i}+v_{j})$. The dimensionless constant $\alpha_{ij,k}$ inside the summations depends on the Fermi velocities,
\begin{eqnarray}
\alpha _{ij,k} \equiv 
\frac{(v_{i}+v_{k})(v_{j}+v_{k})}{2v_{k}(v_{i}+v_{j})}.
\end{eqnarray}
Note that $\alpha_{ij,k}=1$ when $v_{i}=v_{k}$ or $v_{j}=v_{k}$. Finally, 
several short-hand notations are defined,
\begin{eqnarray}
h_{ij} &\equiv& 2f_{ij}+\delta _{ij}c_{ii}, \\
u_{ij}^{0} &\equiv& u_{ii}+u_{jj}, \\
A_{ij}^{p} &\equiv& A_{i{\hat{i}}}+A_{j{\hat{j}}}
+2pA_{i{\hat{j}}},
\end{eqnarray}
where $A=\{c,f,s\}$ and $p=0,\pm 1$.

\end{document}